\documentclass[aip,rsi,reprint,superscriptaddress,showpacs,showkeys,amsmath,amssymb,nofootinbib]{revtex4-1}

\usepackage[caption=false,labelformat=parens]{subfig}
\usepackage{color}
\usepackage{graphicx}
\usepackage{dcolumn}
\usepackage{bm,bbm}
\usepackage{paralist} 
\usepackage{hyperref}
\usepackage[bottom]{footmisc}

\begin{document}

\title{Polarisation in Spin-Echo Experiments: Multi-point and Lock-in Measurements}

\author{Anton Tamt\"{o}gl}
\email{tamtoegl@gmail.com}
\author{Benjamin Davey}
\author{David J. Ward}
\author{Andrew P. Jardine}
\author{John Ellis}
\author{William Allison}
\affiliation{Cavendish Laboratory, J. J. Thompson Avenue, Cambridge, CB3 0HE, United Kingdom}

\date{\today}

\begin{abstract}
Spin-echo instruments are typically used to measure diffusive processes and the dynamics and motion in samples on ps and ns timescales. A key aspect of the spin-echo technique is to determine the polarisation of a particle beam. We present two methods for measuring the spin polarisation in spin-echo experiments. The current method in use is based on taking a number of discrete readings. The implementation of a new method involves continuously rotating the spin and measuring its polarisation after being scattered from the sample. A control system running on a microcontroller is used to perform the spin rotation and to calculate the polarisation of the scattered beam based on a lock-in amplifier. First experimental tests of the method on a helium spin-echo spectrometer, show that it is clearly working and that it has advantages over the discrete approach i.e. it can track changes of the beam properties throughout the experiment. Moreover, we show that real-time numerical simulations can perfectly describe a complex experiment and can be easily used to develop improved experimental methods prior to a first hardware implementation.
\end{abstract}

\keywords{SPIN ECHO EXPERIMENTS, ATOMIC AND MOLECULAR BEAMS, POLARISATION MEASUREMENT, DIFFUSION, LOCK-IN AMPLIFIER}

\maketitle

\section{Introduction}
Spin-echo instruments provide unique information about the dynamics and motion of atoms, molecules and macromolecular objects\cite{Mezei1987,Frick1995,Swenson2001,Richter2005,Alexandrowicz2007,Jardine2009,Hedgeland2009,Fouquet2010,Tamtogl2018}: Upon inelastic scattering from a sample the initial polarisation of a polarised beam is changed. The phase shift of the scattered beam includes important information about the sample dynamics. Hence one of the key aspects of the spin-echo technique is to determine the polarisation of a particle beam. Typically this is done by measuring a couple of discrete points while scanning around the echo condition\cite{DeKieviet1995,Mezei2003}. In this work we present two methods, the above described discrete sampling and a new method based on a continuous spin rotation, in the framework of helium spin-echo spectroscopy.\\
Helium Spin-Echo (HeSE) spectroscopy is a novel technique which combines the surface sensitivity and the inert, completely non-destructive nature of helium atom scattering\cite{Farias1998,Benedek1994,Kraus2013} with the unprecedented energy resolution of the spin-echo method\cite{DeKieviet1995,Gaehler1996,Mezei2003,Fouquet2010}. It allows the study of surface dynamics over time-scales down to sub-picoseconds with atomic resolution\cite{Alexandrowicz2007,Jardine2009}. The apparatus has previously been used to measure diffusion on surfaces\cite{Hedgeland2009,Bahn2017}, surface phonon spectra\cite{Tamtogl2015,Tamtogl2017} and ultra-high resolution potential energy landscapes \cite{Jardine2004}. The polarisation of the He beam is measured by taking a number of discrete points (four) during an echo-scan. We begin by considering this four-point method and note that it suffers from the need for a separate calibration of beam energy and that it also assumes the beam energy is stable throughout the experiment. Starting with a numerical model we show that a new method is capable of monitoring the polarisation continuously. Preliminary experiments using a microcontroller based hardware show clearly that the method works and has advantages over the conventional discrete approach: It requires no initial calibration and adapts to changing conditions.\\

\section{Experimental Setup}
\begin{figure}[htb]
\centering
\includegraphics[width=0.45\textwidth]{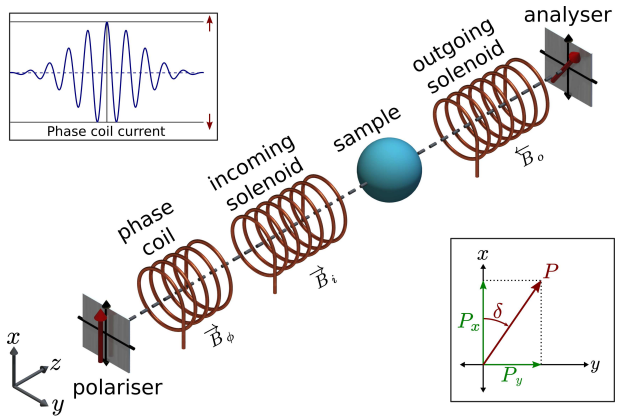}
\caption{A schematic version of the Cambridge spin-echo experiment. A $^3$He beam with the velocity along the $z$-direction is generated and the nuclear spin is polarised along $x$. Before reaching the sample, the beam travels through the magnetic field of the incoming solenoid where the spins perform a Larmor precession in the $xy$ plane. The $^3$He beam is then scattered from the sample and passes thorough the outgoing solenoid where the Larmor precession is effectively unwound with a magnetic field aligned antiparallel to the incoming one. For purely elastic scattering the spin is again aligned along $x$ at the analyser whereas a loss of polarisation is observed in the case of inelastic scattering.\\
The phase coil is included before the incoming solenoid and allows the spin to be rotated by an additional chosen amount. The top left inset illustrates the spin polarisation versus the phase coil current for purely elastic scattering. The phase coil is used to determine both the real and the complex part of an arbitrary spin polarisation in the $xy$ plane (see the inset at the bottom right).}
\label{fig:apparatus}
\end{figure}

In order to perform HeSE measurements, helium-3 ($^3$He) is used which has a non-zero nuclear magnetic moment (spin). The spin is aligned perpendicular to the beam axis, and in the Cambridge instrument scattered of the sample surface in a fixed 44.4$^{\circ}$ source-target-detector geometry. A schematic of the $^3$He spin-echo apparatus is illustrated in \autoref{fig:apparatus} (with the 44.4$^{\circ}$ scattering geometry removed for simplicity) which shows the important components\cite{Fouquet2005}.\\
A nearly monochromatic beam of $^3$He is generated with the velocity along the $z$-direction. The beam passes through a spin polariser, where the nuclear spin is polarised perpendicular to the beamline (along $x$). Before reaching the sample, the beam passes through a magnetic field parallel to the beam axis, which is generated by the incoming solenoid. Thus, the spins perform a Larmor precession in the $xy$ plane where the angle between the $^3$He spin before and after passing the solenoid depends on the time spent in the magnetic field and hence on its velocity. The $^3$He beam is then scattered from the sample and travels through the outgoing solenoid which creates a magnetic field with an amplitude usually equal to that of the incoming coil but now directed anti-parallel to the beam axis. Hence the Larmor precession unwinds the spin so that the spin direction at the end of the outgoing coil is the same as at the beginning of the incoming coil (along the $x$-direction) provided that the scattering event is purely elastic. Finally, the analyser only selects components of the beam with the spin aligned along $x$ after which a detector converts the flux of $^3$He into a count rate.\\
After inelastic or quasi-elastic scattering from the sample surface due to dynamical processes, the signal will only be partially polarised. Consequently, a crucial part of these experiments is to determine the polarisation of the scattered beam.\\

\section{Polarisation measurement}
By measuring how the polarisation changes with increasing current through the solenoids the dynamic properties of the sample can be found. An arbitrary polarisation $P$ of the spin in the $xy$-plane can be conveniently written as a complex number,
\begin{equation}
P = P_x + \mathbbm{i} P_y = |P| \cos{\delta} + \mathbbm{i} |P| \sin{\delta} \; ,
\end{equation}
where $P_x = \operatorname{\mathbb{R}e} ( P )$ and $P_y = \operatorname{\mathbb{I}m} ( P )$ (see also the illustration in the inset of \autoref{fig:apparatus}). In principle it is not essential to measure both $P_x$ and $P_y$ in every experiment. E.g. a measurement of $P_x$ gives the real part of the intermediate scattering function $ISF(\Delta K,t)$. However, some inelastic processes such as phonon events require the measurement of both components of the polarisation\cite{Jardine2009}.\\
Moreover, even without a magnetic field in the solenoids ($B_i = B_o = 0$) the polarisation may not be entirely real due to practical hardware implementations. These include the misalignment of magnetic elements, small differences between the two coils and their driving circuits as well as external magnetic fields penetrating the beam path. The issue is illustrated in \autoref{fig:FourPoint} which shows schematically a line corresponding to a polarisation measurement under ideal conditions (black, dashed) and a realistic elastic measurement (blue, solid). Hence to obtain the maximum amplitude of the polarisation (due to practical limitations of the apparatus) and the complete information about the underlying dynamical processes both the real and the imaginary part of the polarisation need to be measured.\\
In principle the simplest way of measuring the complex polarisation is to rotate the analyser (in \autoref{fig:apparatus}), by $\pi / 2$ which allows independent measurements of the polarisation along the $x$ and the $y$-axis: $P_x$ and $P_y$. However, mechanical rotations of magnetic devices on the instrument are quite complicated and in practice there are simpler solutions. The most convenient approach is to add an additional ``phase coil'' in the region before the incoming solenoid (see \autoref{fig:apparatus}). The phase coil rotates the spin by a chosen amount depending on the magnetic field $B_{\phi}$. Thus the ingoing spin vector can be rotated with respect to the outgoing one by a certain amount which is determined by the current through the phase coil. By measuring a number of discrete points within the period of one full spin rotation the exact polarisation of the spin can be measured. In neutron spin-echo spectroscopy this is typically termed as measurement of the echo-group or as so-called echo-scan\cite{Mezei2003,Farago2003,Ohl2012}. The process is illustrated for four points in \autoref{fig:FourPoint} and will be described in the following.

\subsection{The 4-Point Measurement}
\label{sec:fourpoint}
\begin{figure}[htb]
\centering
\includegraphics[width=0.48\textwidth]{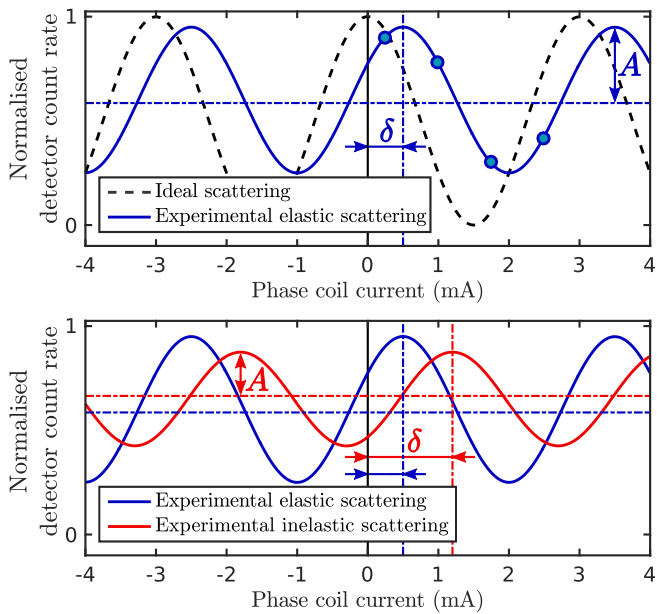}
\caption{Illustration of the 4-point measurement (colour online). Upper panel: The dashed curve shows the variation in the polarised intensity with phase coil current assuming an ideal scattering experiment from a static sample. The blue solid curve represents scattering from a static sample as it appears in the experiment. Due to non-idealities the polarisation of the beam is not perfect and a background as well as a phase shift with respect to the dashed curve occur. In order to determine the polarisation, four points are measured over one oscillation which is illustrated by the four dots on the blue curve. By fitting  \autoref{equ:phasecoil} to these points the amplitude $A$ and the phase shift $\delta$ are found which is then used to calculate the real and the imaginary part of the polarisation.\\
Lower panel: The red solid line illustrates an inelastic scattering experiment where motion gives rise to an additional phase shift and a change in the amplitude of the signal with respect to scattering from a static sample (blue curve).}
\label{fig:FourPoint}
\end{figure}

As described in the previous section, the complex polarisation of the beam can be measured via an additional spin rotation in the phase coil. With a current, $I$, passing through the phase-coil, a spin is rotated through an angle $\Omega_0 I$ where the rotation constant $\Omega_0 = \gamma m B_{eff} \lambda_0 / \hbar$, depends on the mean wavelength of the particles in the beam, $\lambda_0$, the particle mass and the field integral along the beam path $B_{eff} = \int B \: \mathrm{d}z /I $\cite{Jardine2009,Alexandrowicz2007,Jones2016}. The signal resulting from a fixed polariser in the case of a beam with a Gaussian spread of wavelengths is derived in the Supplementary Information and is
\begin{equation}
P_x (I) = A \; \mathrm{e} ^{(- I^2 / 2 \sigma^2 )} \; \cos \left[ \Omega_0 (I - \delta) \right] +  C  \; .
\label{equ:phasecoil1}
\end{equation}
In practice the polarisation is measured via the $^3$He atoms that arrive at the detector where the flux is converted into a count rate
\begin{equation}
n_{det}(I) = A \; \mathrm{e} ^{(- I^2 / 2 \sigma^2 )} \; \cos \left[ \Omega_0 (I - \delta) \right] +  \underbrace{C_1 + C_2}_{C}  \; .
\label{equ:phasecoil}
\end{equation}
The count rate $n_{det}$ in the detector follows \eqref{equ:phasecoil} as the current in the phase coil ($I$) is changed (see \autoref{fig:FourPoint}). The $\cos ( \Omega_0 I )$ term describes the rotation of the spin and is determined by the mean wavelength of the particles (via $\Omega_0$) and the current $I$ through the phase coil. The phase shift $\delta$ in \eqref{equ:phasecoil} accounts for non-idealities  within the ingoing and outgoing spin rotation as discussed above: It describes the effect that the polarisation is not entirely real for $I=0$ as illustrated in the upper panel of \autoref{fig:FourPoint}. The spread of wavelengths in the beam is manifest through the Gaussian pre-factor in \eqref{equ:phasecoil}, of width $\sigma = \hbar / \gamma m B_{eff} \sigma_{\lambda}$, where $\sigma_{\lambda}$ is the corresponding spread in wavelengths. 
Finally, the offset $C$ in \eqref{equ:phasecoil} is composed of a background $C_1$ in the detector and an offset $C_2$ which is due to the fact that the polarisation of the $^3$He beam and the selection of a single polarisation by the analyser cannot be perfect in reality.\\
The properties of the beam $\Omega_0$ and $\sigma$ in \eqref{equ:phasecoil} have typically been determined beforehand. When there is little time variation of $\Omega_0$ and $\sigma$ the beam properties may be determined in an infrequent precise measurement over many oscillations. The process involves a measurement of the polarisation for a static sample while scanning the current in the incoming solenoid with the outgoing solenoid current held at zero. At present, the measurement takes about 20 min to perform and is usually done each time the beam is adjusted\cite{Ward2013}. The polarised intensity is then measured for a number of different phase coil currents in a single period of the oscillation, as shown by the example blue markers in \autoref{fig:FourPoint} which is then used to calculate the complex polarisation of the beam.\\
Hence if the properties of the beam are known, there are three remaining variables in \eqref{equ:phasecoil}: $A$, $C$ and $\delta$. These variables can be determined by measuring at three different phase coil currents $I$ over one oscillation period and solving the resulting set of linear equations. In practice we measure 4 points as described above giving rise to an over constrained set of linear equations which is solved using a least squared optimisation. The measurement of four points presents a balance of accurate results and high throughput.\\
Finally after all variables have been determined, \eqref{equ:phasecoil} can be used to calculate the real and the complex part, ${\mathbb{R}e} ( P )$ and $\operatorname{\mathbb{I}m} ( P )$ of the spin polarisation. Any changes of the polarisation due to dynamical processes can be determined using this principle. In general, quasi-elastic and inelastic scattering off the sample gives rise to a change of the phase $\delta$ and amplitude $A$ of the oscillation with respect to an elastic scattering event, which is illustrated in the lower panel of \autoref{fig:FourPoint}.

\subsection{The Spin-Rotator Method}
The conventional method of measuring the polarisation, the four point method described in the previous section, involves passing discrete currents through the phase coil, each rotating the direction of the spin by a different amount and measuring the scattered signal at the detector. Even though this method has proven to be perfectly suitable for most measurements it exhibits a couple of disadvantages. Firstly, the calibration routine used at the start is relatively time consuming ($\approx 20$ min). Then four data points are used to find three parameters which means the system is susceptible to noise, e.g. fluctuations in the detector. Finally it is assumed that the value of $\sigma$ and $\Omega_0$ remain constant throughout the experiment. In the following we present a new method of finding the polarisation which does not require initial calibration, is less susceptible to noise and will adjust to changing values of $\sigma$ and $\Omega_0$.\\
The proposed new method involves continuously rotating the polarisation of the incoming beam instead of rotating it by separate discrete amounts. In doing so, the polarised component will cause an oscillating signal in the detector where the frequency of the oscillation equals the frequency of the spin rotation. Any remaining unpolarised component will give rise to a constant D.C. level in the detector.\\
Hence a lock-in amplifier with the same frequency as the spin rotation can be used to extract the polarised component of the signal (see section S3.A in the supplementary information and \cite{Caldwell1977} for a similar concept in the context of polarisation measurements in optical systems). The unpolarised component can be found by running the detector signal through a low pass filter. Using both components the magnitude of the polarisation and the phase of the spin can be calculated. The method has the advantage of requiring no initial calibration while also taking data continuously.\\
\begin{figure}[htb]
\centering
\includegraphics[width=0.48\textwidth]{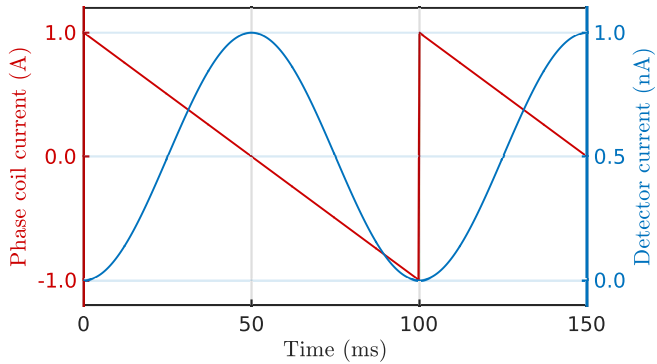}
\caption{Modelled response of the detector signal for a sawtooth current through the phase coil. The sawtooth current in the phase coil results in a sinusoidal current output from the detector. This is based on an idealised model where no delay due to the time-of-flight of the $^3$He atoms and phase shifts due to scattering off the sample have been included.}
\label{fig:figure3}
\end{figure}
In order to rotate the initial spin direction continuously, the current in the phase coil needs to be linearly increased. To obtain this effect with a finite current, the current is first increased linearly and then quickly dropped back to the starting current (a sawtooth wave). If the amplitude is chosen correctly, at the start of each period the $^3$He beam will have the same polarisation direction as at the end of the previous period. \autoref{fig:figure3} illustrates that this will create a smooth sine wave at the detector, based on a numerical simulation with an idealised model of the system (which will be discussed below).\\
If the amplitude is incorrect, there will be a sudden jump in the signal at the end of each period. The correct amplitude is ensured by controlling the amplitude of the modulating sawtooth with feedback from the detector output. \autoref{fig:blockdiagram} shows a block diagram of the new method for the polarisation measurement with the feedback loop in red. The output from the detector passes through a lock-in amplifier with a reference signal at $2\omega$ which controls the amplitude of the sawtooth. Starting with an arbitrary initial amplitude of the sawtooth current in the phase coil, the system should tend towards a stable position where the phase coil amplitude corresponds to one complete rotation of the $^3$He spin. We will discuss this in more depth on the basis of a numerical model of the system below.
\begin{figure}[htb]
\centering
 \includegraphics[width=0.48\textwidth]{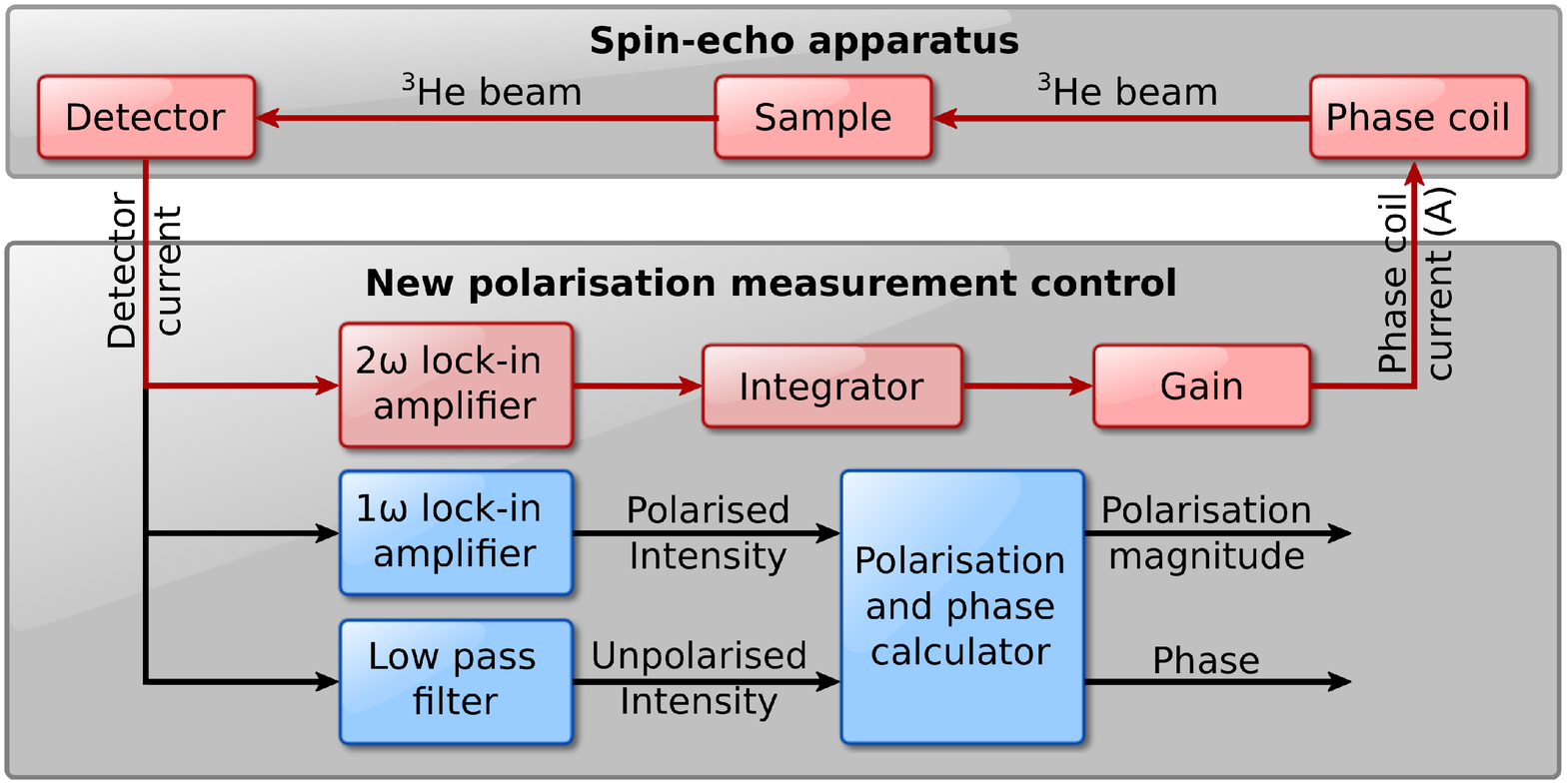}
 \caption{(Colour online) Block diagram of the system to measure the polarisation of the beam. The top block represents the spin-echo machine, the bottom block the new control system. The feedback loop which controls the amplitude of the sawtooth current in the phase coil is shown in red. A $1\omega$ lock-in amplifier and a low pass filter provide the polarised and unpolarised components of the beam, respectively, which allows the magnitude of the polarisation and the phase to be calculated. For the individual components of the block diagram please refer to S3 in the supplementary information.}
 \label{fig:blockdiagram}
\end{figure}

\section{Numerical Model}
A numerical model allows the behaviour of the system to be tested. Starting with an idealised system to check the feasibility of the method it permits the subsequent addition of imperfections of a realistic system step by step so that the effect of each can be found. Simulink (a graphical programming language for modelling dynamic systems) was chosen to model the system as it enables rapid production of a simple model which can then easily be modified with more realistic aspects.

\subsection{An Idealised Numerical Model}
\begin{figure}[htb]
\centering
\includegraphics[width=0.48\textwidth]{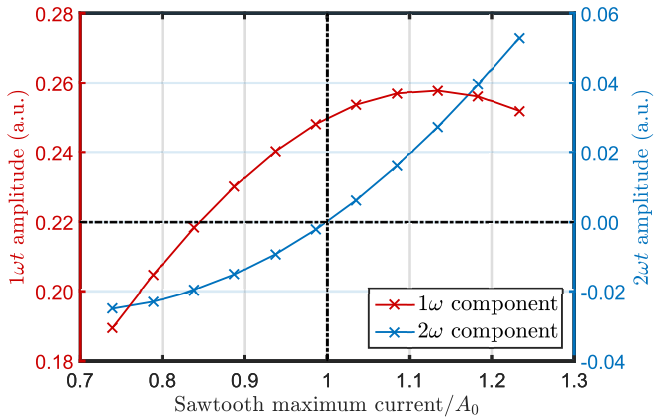}
 \caption{Effect of changing the sawtooth maximum amplitude on the amplitude of the different frequency components. The $2\omega$ signal goes to zero at the perfect amplitude $A_0$ (vertical dash-dotted line) and has a different sign on each side, allowing negative feedback to be used to control the system. The $1\omega$ signal has a maximum at a phase coil current slightly greater than the perfect current, meaning that the current needs to be tightly controlled.}
 \label{fig:perfectampcomponents}
\end{figure}

The behaviour of the aforementioned new method of polarisation measurements was first modelled in Simulink using the components shown in \autoref{fig:blockdiagram}. We have already seen in \autoref{fig:figure3} that the right choice of the sawtooth amplitude in the phase coil gives rise to a sinusoidal current output from the detector. However, the system should find the right amplitude starting from an arbitrary initial value. Therefore a feedback loop (red parts in \autoref{fig:blockdiagram}) consisting of a $2\omega$ lock-in amplifier, an integrator and the feedback loop gain is used. The spin-echo machine and the current output from the detector is modelled to follow equation \eqref{equ:phasecoil}.\\
\autoref{fig:perfectampcomponents} shows that a non-perfect amplitude introduces higher Fourier components to the signal. At the perfect amplitude, the signal at twice the sawtooth frequency (2$\omega$) goes to zero as expected. To ensure that the amplitude is kept at the correct value, a system of negative feedback can be used as the $2\omega$ signal changes sign at the perfect amplitude (vertical dash-dotted line in \autoref{fig:perfectampcomponents}).\\
The feedback loop is created by integrating the $2\omega$ component to give it an infinite DC gain and to act as the dominant pole in the loop\cite{Pippard1985}. The integrated signal is multiplied by an adjustable constant, known as the feedback gain and then multiplied by a unit amplitude sawtooth wave to supply the current in the phase coil. Finally, a $1\omega$ lock-in amplifier and a low pass filter provide the polarised and unpolarised components of the signal, respectively, which are used to calculate the magnitude of the polarisation and the phase (see \autoref{fig:blockdiagram}).\\

\subsection{A Realistic Numerical Model}
The model described above assumes an ideal system which corresponds instantaneously to any changes and is free from any background noise. As a more realistic model, three sources of imperfection were added: 
\begin{compactitem}
\item Background noise.
\item The continuous change of current in the phase coil compared to the ideal instantaneous jump: Due to the inductance of the phase coil the rapid change of current at the end of the sawtooth period (\autoref{fig:figure3}) will not be instantaneous. Instead the current will change continuously and during this time, the spin polarisation is rotated by a full cycle very quickly leading to a short blip in the detector current. This blip will give rise to frequency components across the whole spectrum but it will not affect the stability of the system as shown by tests with the realistic model and also in the measurements later.
\item The response function of the apparatus consisting of the time-of-flight of the $^3$He atoms and the response of the detector. 
\end{compactitem}
The last point is also interesting from an experimental point of view and has not been determined previously. Hence we will discuss it in the next section. The setup of the Simulink model itself can be found in the Supplementary Information.

\subsection{Apparatus Response Function}
\begin{figure}[htb]
\centering
 \includegraphics[width=0.48\textwidth]{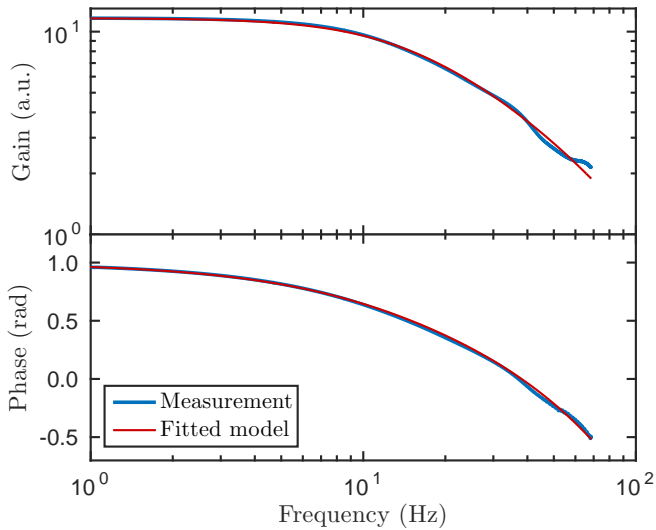}
 \caption{(Colour online) Bode plot of the response function for the Cambridge $^3$He spin-echo machine. Plotted in blue is the experimentally measured gain and phase. The red line shows a fit of the gain and phase based on \autoref{equ:transferfunction}. The fitted parameters are listed in \autoref{tab:responseparameters}.}
 \label{fig:bodeplot}
\end{figure}

After the field in the phase coil is changed, the $^3$He atom must travel the length from the phase coil to the detector where it is ionised and then converted to a count rate, meaning the response is not instantaneous. Since the $^3$He atoms do not interact with each other, a linear response function represents a good model of the machine.\\
The frequency response of the apparatus was determined by applying a square wave of fixed frequency to the phase-coil. The current in the phase-coil and the detector count rate were both measured and Fourier transforms of both were taken, they were then used to calculate the gain and phase shift as a function of frequency. The response determined from the upward transition of the square wave was identical to that from the downward transition, confirming the linearity of the system.  An average over several cycles leads to the frequency response shown in \autoref{fig:bodeplot} (blue lines).\\
A transfer function of the form, \autoref{equ:transferfunction}, was fitted to the measured data where $s$ is the Laplace transform variable and $\tau$ is a time delay due to the time-of-flight.
\begin{equation}
H(s)=\frac{DCgain\cdot \operatorname{e}^{-s\tau}}{\left(\frac{s}{2\pi f_{1}}+1\right) \left( \frac{s}{2\pi f_{2}}+1\right)} \; .
\label{equ:transferfunction}
\end{equation}
\autoref{equ:transferfunction} has a gain $G$ and phase $\phi $ according to
\begin{equation}
\begin{aligned}
G(f) &= \operatorname{abs} \left[ \frac{DCgain}{\left(\frac{\mathbbm{i}f}{f_{1}}+1\right) \left( \frac{\mathbbm{i}f}{f_{2}}+1\right)} \right] \; , \\
\phi (f) &= \tan ^{-1} \left\{ \frac{\operatorname{\mathbb{R}e} \left[ \frac{DCgain\cdot \operatorname{e}^{-\mathbbm{i}2\pi f\tau}}{\left(\frac{\mathbbm{i}f}{f_{1}}+1\right) \left( \frac{\mathbbm{i}f}{f_{2}}+1\right)} \right] }{\operatorname{\mathbb{I}m} \left[ \frac{DCgain\cdot e^{-\mathbbm{i}2\pi f\tau}}{\left(\frac{\mathbbm{i}f}{f_{1}}+1\right) \left( \frac{\mathbbm{i}f}{f_{2}}+1\right)} \right] } \right\}.
\end{aligned}
\label{equ:gainphase}
\end{equation}
The gain was fitted first since it is most sensitive to changes in $f_1$, $f_2$ and the $DCgain$. Once a fit for that had been found, the value of $\tau$ was fitted for the phase. The values of the fitted parameters are summarised in \autoref{tab:responseparameters} and the fitted response, plotted in red in \autoref{fig:bodeplot}, shows a good agreement between the measurement and the model.\\
\begin{table}[htb]
\centering
\begin{tabular}{ c | c }
    Parameter & Value \\ \hline
    $DCgain$ & $-11.66\pm 0.03$ \\
      $f_1$ & $(14.61\pm 0.03)$ Hz  \\
     $f_2$ & $(84.6 \pm 0.6)$ Hz\\
     $\tau$ & $(6.328 \pm 0.005)$ ms \\
\end{tabular}
\caption{Best-fit parameters which describe the measured apparatus response function according to \autoref{equ:transferfunction}.}
\label{tab:responseparameters}
\end{table}
The parameter $\tau$ is expected to be the time delay due to the time-of-flight of the $^3$He atom from the phase coil to the detector. The length of this section of the machine is known from\cite{Jardine2009} ($\approx 4.7$ m) which would give a time-of-flight of $\approx 6.3$ ms for an 8 meV $^3$He beam. This value agrees very well with the value obtained from the measurement (\autoref{tab:responseparameters}), indicating that the time-of-flight is indeed the dominant contribution to the delay. Note that the gain obtained from the measurement will change from experiment to experiment since it depends on properties such as the beam intensity and the reflectivity of the sample.\\
The poles of the response function correspond to time constants of approximately 10 ms and 2 ms; these are due to the diffusion and ionisation time scales within the detector respectively. It is important to bear these in mind when choosing the frequency at which the system should run at. Above these poles, phase changes occur which may change negative feedback to positive feedback and the gain starts to decrease, reducing the measurable signal, but as the frequency is lowered, the system will respond slower to any changes. In the following experiments a frequency of 5 Hz was used.\\

\subsection{Closed Loop Response of the Realistic Model}
\begin{figure}
\centering
\includegraphics[width=0.48\textwidth]{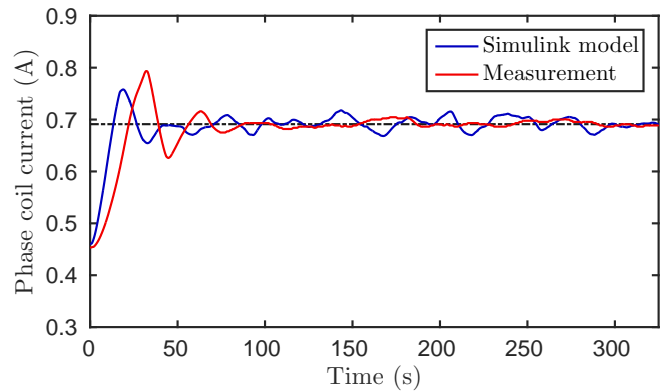}
\caption{(Colour online) Comparison between the Simulik model and the hardware version. Blue line: Closed loop response of the realistic numerical model shown by monitoring the maximum amplitude of the current through the modelled phase coil. The dash-dotted horizontal line illustrates the theoretical perfect amplitude. The system follows a typical negative feedback behaviour, tending to the perfect value, overshooting and oscillating around it with decaying amplitude.\\
Red line: The behaviour of the hardware version of the polarisation control system tracked by measuring the maximum current through the ``real'' phase coil. The system shows similar behaviour to the realistic model.}
\label{fig:imperfectresponse}
\end{figure}
The stability of the realistic numerical model was tested by tracking the amplitude of the sawtooth throughout the experiment. Starting with an arbitrary initial amplitude the system should then tend towards the ideal amplitude which corresponds to exactly one rotation of the spin. Therefore the maximum amplitude of the current in the phase coil is monitored versus time, which is shown in \autoref{fig:imperfectresponse}. The horizontal dash-dotted line illustrates the theoretical perfect amplitude based on the beam energy and solenoid properties.\\
The system follows a typical negative feedback behaviour: The amplitude tends to the theoretical correct value, overshoots slightly before oscillating with decaying amplitude around the perfect value. The time delay and response function for the detector had no noticeable effect on whether the system was stable for a range of lock-in amplifier time constants and feedback gains. However, the realistic system tends to oscillate around the constant value with a higher frequency and higher amplitude compared to the idealised model. Size/frequency of these oscillations can be changed by optimising the parameters of the feedback loop (gain, time constant)\cite{Taylor1994}.

\section{Hardware Implementation}
\begin{figure}
\centering
  \centering
  \includegraphics[width=0.48\textwidth]{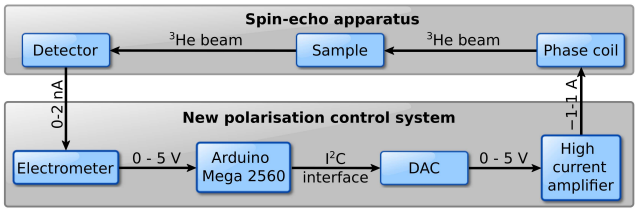}
  \caption{Block diagram of the hardware to provide feedback regulation in the new polarisation control system.}
 \label{fig:blockdiagram2}
\end{figure}

Since the model showed that it is theoretically possible to measure the polarisation with the new method a hardware version was implemented which can be connected to the spin-echo machine. In order to do this, the Simulink program which controls the feedback and calculates the polarisation needs to run in real time. It also needs to drive the current through the phase coil and read the current from the detector back in.\\
The control system was chosen to run on a microcontroller, an Arduino Mega 2560. It has enough internal memory to store the required programme, can send and receive data through a high speed I$^2$C interface and can send data to a PC for logging purposes. For the typical beam energies of the apparatus ($6-12$ meV) a maximum current of 0.9 A through the phase coil is required to guarantee a full spin rotation at the highest energy beam. Therefore the Arduino is connected to a 12 bit digital to analogue converter (DAC, MCP4725) via the I$^2$C interface. The $0-5$ V of the DAC are then used to provide the phase coil current via a high current amplifier (see Supplementary Information for the circuit diagrams). Thus a range of $-1$ A to $+1$ A through the phase coil is achieved, which meets the standard conditions and also gives some space for overshoots of the sawtooth wave.\\
The built-in analogue input on the Arduino was used to read the detector current. Since the output from the detector varies between $10^{-7}$ and $10^{-13}$ A, a simple electrometer converts the current into a $0-5$ V output. For a first test it was decided to focus on the middle range (1 nA range). A block diagram of the final setup is shown in \autoref{fig:blockdiagram2}. 

\section{Experimental Results}
To test the apparatus on the spin echo machine a graphene/Ni(111) surface was used\cite{Tamtogl2015} with the scattering slightly off specular. This arrangement was chosen since it gives a detector current in the desired range as well as a partly polarised scattered beam. The process was monitored by recording the current though the phase coil and the current in the detector, simultaneously, via an oscilloscope (see Supplementary Information for more details).\\
\begin{figure}
\centering
  \centering
  \includegraphics[width=0.48\textwidth]{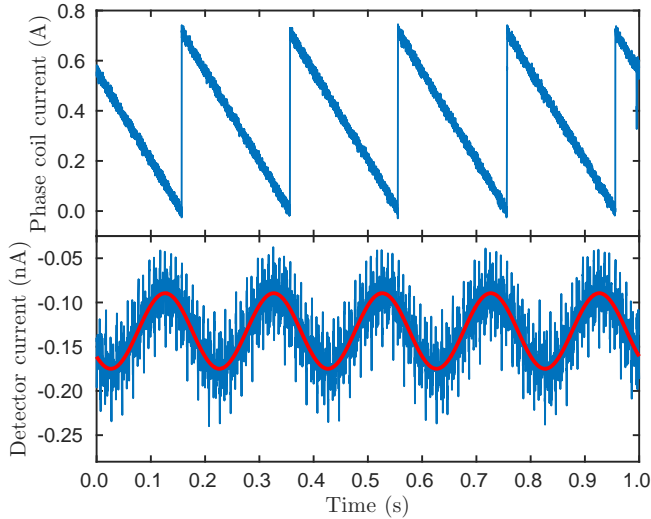}
     \caption{(Colour online) Phase coil and detector currents for the hardware system. Top panel: The current in the phase coil follows a curved sawtooth. Bottom panel: current received from the detector in blue and a fitted sine wave in red. The sinusoidal shape of the current from the detector indicates that the circuit has successfully locked in.}
        \label{fig:inputoutputreal}
\end{figure}
After the system has reached stability the current measured in the phase coil (top panel of \autoref{fig:inputoutputreal}) shows the expected sawtooth shape and the current from the detector  (bottom panel of \autoref{fig:inputoutputreal}) shows a definite sine wave. The red curve in \autoref{fig:inputoutputreal} is a sine fit to the detector current. The frequency of the sine wave is 5 Hz indicating that the system has successfully locked in and that the concept is working.\\
The measured signal exhibits still a large fraction of noise which is mainly due to the simple setup of the electric circuits for reading in the detector current. Nevertheless, the first test shows that the spin-rotator method is perfectly working and we will discuss its advantages over the 4-point method at the end of this section. Furthermore, the test also shows that a lock-in amplifier can be easily implemented and run on a microcontroller, similar in spirit to previous works\cite{Gonzalez2007,Schultz2016}.\\
The stability of the system can be tested in the same manner as the numerical model. Therefore, the maximum amplitude of the phase coil current is plotted in red in \autoref{fig:imperfectresponse}. After the system has been switched on with an initial amplitude that is much smaller than the perfect amplitude it again overshoots before it oscillates slightly around the perfect amplitude. The amplitude has stabilised within less than a minute. The comparison with the numerical model (\autoref{fig:imperfectresponse}) shows that the similarity between the real system and the model is remarkable.\\
\begin{figure}
\centering
  \centering
  \includegraphics[width=0.48\textwidth]{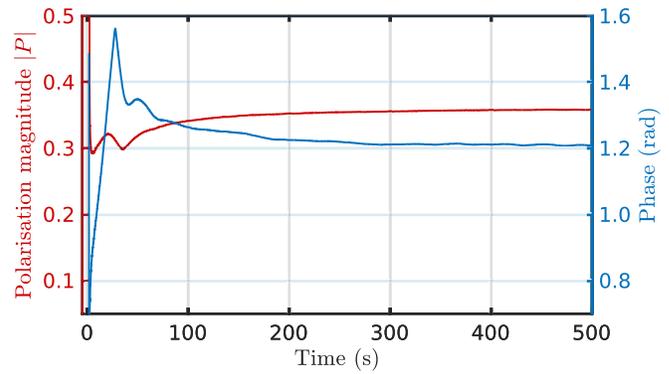}
  \caption{Measurement of the polarisation and phase as calculated by the microcontroller versus time. Starting with t=0 the system is switched on and it takes some time until the phase coil current reaches the perfect amplitude. Once the control system has stabilised both polarisation and phase tend towards a constant value.}
  \label{fig:polandphasereal}
\end{figure}
These experiments had all shown the system to be stable so the next test was to try taking a measurement of the beam polarisation. The polarisation and phase can be seen in \autoref{fig:polandphasereal} to tend to a stable polarisation of $0.3582\pm0.0004$. Note that the maximum polarisation of the $^3$He beam is $0.6$ due to the polariser and transmission properties of the apparatus\cite{McIntosh2016}. Hence this corresponds to a polarisation of roughly 60\% for the scattered $^3$He beam. The uncertainty of the measured polarisation shows a factor of three improvement with the new method compared to the 4-point measurement. The measurement of the phase settled on a constant value with fluctuations of 0.0018 rad, again the four-point method did not measure the phase to an accuracy of 0.004 rad. On the other hand, a direct comparison with the four-point method is difficult to make, since the system started from an arbitrary initial value and the measurement was taken after a longer time.\\
\begin{figure}
\centering
  \centering
  \includegraphics[width=0.48\textwidth]{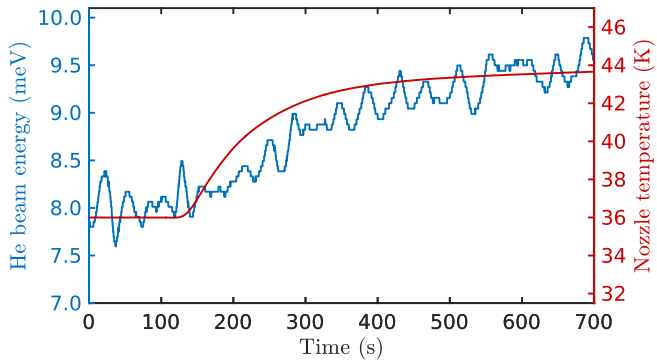}
  \caption{(Colour online) The new method can track changes of the beam energy which is not possible with the 4-point method. Blue curve: Sawtooth amplitude which controls the current through the phase coil. Red curve: Temperature of the helium beam nozzle which is proportional to the beam energy. With increasing beam energy the amplitude of the sawtooth needs to increase in order to maintain a full spin rotation in the phase coil. The amplitude clearly follows the trend of the nozzle temperature.}
  \label{fig:changepol}
\end{figure}
The ability to monitor polarisation continuously is a significant advantage of the spin-rotator method. As mentioned in \autoref{sec:fourpoint}, the 4-point measurement assumes a constant beam profile throughout a run of experiments. The behaviour of the new polarisation control system with respect to this aspect can be tested by actively changing the beam properties, i.e. by changing the helium beam energy. The beam energy $E_0$ is determined by the temperature of the nozzle $T_N$, which produces the atomic beam, via $E_0 = 5/2 k_B T_N$. Here, $k_B$ is the Boltzmann constant\cite{Farias1998}. Note however, that the nozzle temperature cannot be used for an exact measurement of the beam energy since there might be small differences between the diode-based temperature measurement and the actual temperature inside the nozzle\cite{Lechner2013}.\\
The blue curve in \autoref{fig:changepol} shows the beam energy calculated from the output voltage of the microcontroller. The red curve in \autoref{fig:changepol} displays the nozzle temperature which is proportional to the $^3$He beam energy. Starting with a nozzle temperature of 36 K, after about 150 s the nozzle temperature starts to increase, from about 8 to 9.5 meV. The amplitude of the sawtooth has to increase with increasing beam energy in order to maintain a full spin rotation in the phase coil. While there are still small oscillations on the beam energy determined from the control system, it definitely follows the general trend of the nozzle temperature. Hence the new system is stable regardless of the beam energy and can track changes of the beam properties throughout the experiment.\\
In cases where the beam polarisation is negligible for an extended period it is likely that the simple integrator in the $2 \omega$ feedback loop (\autoref{fig:blockdiagram}) will cause the amplitude of the current in the phase-coil to drift. The feedback loop will restore the correct amplitude as soon as the polarisation returns. However, if the drift is problematic it would be straightforward to add a sample/hold element controlled by the polarisation magnitude so that the feedback loop is inhibited and the phase-coil amplitude fixed, whenever the polarisation falls below a certain pre-set value.

\section{Summary and Conclusion}
In summary, we have described two methods of measuring the polarisation in spin-echo experiments: The four-point method, which is based on taking discrete points along an echo scan and a new, so called spin-rotator method. Numerical simulations and a novel hardware implementation have been used to develop the new method which involves rotating the spin and measuring its polarisation after scattering continuously. A control system implemented on a microcontroller is used to control the spin rotation and to calculate the polarisation of the scattered beam, based on a lock-in amplifier.\\
First experimental tests of the new method on a helium spin-echo apparatus show clearly that the method is working. While the first hardware implementation of the spin-rotator method may still require a number of improvements in terms of noise reduction and speed, the first tests show that it has advantages over the conventional discrete approach. The four-point method, which is currently used, requires a preceding calibration at the start of an experiment and assumes a constant beam profile. The spin-rotator method, on the other hand, does not assume a constant beam profile and can track changes of the beam properties throughout the experiment.\\
Moreover, our numerical model shows, that real-time numerical simulations are capable of accurately describing a complex experiment. They can be used to test the behaviour and viability prior to first experiments. Finally, our tests show that a lock-in amplifier can be simply implemented on a cheap microcontroller and perform the tasks of a bench-top lock-in amplifier.

\section*{Supplementary Material}
The setup of the numerical model in Simulink and the circuit diagrams of the hardware implementation can be found in the supplementary material.

\section*{Acknowledgement}
One of us (A.T.) acknowledges financial support provided by the FWF (Austrian Science Fund) within the project J3479-N20.

\bibliography{literature}

\begin{thebibliography}{31}%
\makeatletter
\providecommand \@ifxundefined [1]{%
 \@ifx{#1\undefined}
}%
\providecommand \@ifnum [1]{%
 \ifnum #1\expandafter \@firstoftwo
 \else \expandafter \@secondoftwo
 \fi
}%
\providecommand \@ifx [1]{%
 \ifx #1\expandafter \@firstoftwo
 \else \expandafter \@secondoftwo
 \fi
}%
\providecommand \natexlab [1]{#1}%
\providecommand \enquote  [1]{``#1''}%
\providecommand \bibnamefont  [1]{#1}%
\providecommand \bibfnamefont [1]{#1}%
\providecommand \citenamefont [1]{#1}%
\providecommand \href@noop [0]{\@secondoftwo}%
\providecommand \href [0]{\begingroup \@sanitize@url \@href}%
\providecommand \@href[1]{\@@startlink{#1}\@@href}%
\providecommand \@@href[1]{\endgroup#1\@@endlink}%
\providecommand \@sanitize@url [0]{\catcode `\\12\catcode `\$12\catcode
  `\&12\catcode `\#12\catcode `\^12\catcode `\_12\catcode `\%12\relax}%
\providecommand \@@startlink[1]{}%
\providecommand \@@endlink[0]{}%
\providecommand \url  [0]{\begingroup\@sanitize@url \@url }%
\providecommand \@url [1]{\endgroup\@href {#1}{\urlprefix }}%
\providecommand \urlprefix  [0]{URL }%
\providecommand \Eprint [0]{\href }%
\providecommand \doibase [0]{http://dx.doi.org/}%
\providecommand \selectlanguage [0]{\@gobble}%
\providecommand \bibinfo  [0]{\@secondoftwo}%
\providecommand \bibfield  [0]{\@secondoftwo}%
\providecommand \translation [1]{[#1]}%
\providecommand \BibitemOpen [0]{}%
\providecommand \bibitemStop [0]{}%
\providecommand \bibitemNoStop [0]{.\EOS\space}%
\providecommand \EOS [0]{\spacefactor3000\relax}%
\providecommand \BibitemShut  [1]{\csname bibitem#1\endcsname}%
\let\auto@bib@innerbib\@empty
\bibitem [{\citenamefont {Mezei}, \citenamefont {Knaak},\ and\ \citenamefont
  {Farago}(1987)}]{Mezei1987}%
  \BibitemOpen
  \bibfield  {author} {\bibinfo {author} {\bibfnamefont {F.}~\bibnamefont
  {Mezei}}, \bibinfo {author} {\bibfnamefont {W.}~\bibnamefont {Knaak}}, \ and\
  \bibinfo {author} {\bibfnamefont {B.}~\bibnamefont {Farago}},\ }\href
  {\doibase 10.1103/PhysRevLett.58.571} {\bibfield  {journal} {\bibinfo
  {journal} {Phys. Rev. Lett.}\ }\textbf {\bibinfo {volume} {58}},\ \bibinfo
  {pages} {571} (\bibinfo {year} {1987})}\BibitemShut {NoStop}%
\bibitem [{\citenamefont {Frick}\ and\ \citenamefont
  {Richter}(1995)}]{Frick1995}%
  \BibitemOpen
  \bibfield  {author} {\bibinfo {author} {\bibfnamefont {B.}~\bibnamefont
  {Frick}}\ and\ \bibinfo {author} {\bibfnamefont {D.}~\bibnamefont
  {Richter}},\ }\href {http://science.sciencemag.org/content/267/5206/1939}
  {\bibfield  {journal} {\bibinfo  {journal} {Science}\ }\textbf {\bibinfo
  {volume} {267}},\ \bibinfo {pages} {1939} (\bibinfo {year}
  {1995})}\BibitemShut {NoStop}%
\bibitem [{\citenamefont {Swenson}, \citenamefont {Bergman},\ and\
  \citenamefont {Longeville}(2001)}]{Swenson2001}%
  \BibitemOpen
  \bibfield  {author} {\bibinfo {author} {\bibfnamefont {J.}~\bibnamefont
  {Swenson}}, \bibinfo {author} {\bibfnamefont {R.}~\bibnamefont {Bergman}}, \
  and\ \bibinfo {author} {\bibfnamefont {S.}~\bibnamefont {Longeville}},\
  }\href {\doibase 10.1063/1.1420728} {\bibfield  {journal} {\bibinfo
  {journal} {J. Chem. Phys}\ }\textbf {\bibinfo {volume} {115}},\ \bibinfo
  {pages} {11299} (\bibinfo {year} {2001})}\BibitemShut {NoStop}%
\bibitem [{\citenamefont {Richter}\ \emph {et~al.}(2005)\citenamefont
  {Richter}, \citenamefont {Monkenbusch}, \citenamefont {Arbe},\ and\
  \citenamefont {Colmenero}}]{Richter2005}%
  \BibitemOpen
  \bibfield  {author} {\bibinfo {author} {\bibfnamefont {D.}~\bibnamefont
  {Richter}}, \bibinfo {author} {\bibfnamefont {M.}~\bibnamefont
  {Monkenbusch}}, \bibinfo {author} {\bibfnamefont {A.}~\bibnamefont {Arbe}}, \
  and\ \bibinfo {author} {\bibfnamefont {J.}~\bibnamefont {Colmenero}},\ }\href
  {\doibase 10.1007/b106578} {\emph {\bibinfo {title} {Neutron Spin Echo in
  Polymer Systems}}},\ \bibinfo {series} {Advances in Polymer Science}, Vol.\
  \bibinfo {volume} {174}\ (\bibinfo  {publisher} {Springer Berlin
  Heidelberg},\ \bibinfo {address} {Berlin, Heidelberg},\ \bibinfo {year}
  {2005})\ pp.\ \bibinfo {pages} {1--221}\BibitemShut {NoStop}%
\bibitem [{\citenamefont {Alexandrowicz}\ and\ \citenamefont
  {Jardine}(2007)}]{Alexandrowicz2007}%
  \BibitemOpen
  \bibfield  {author} {\bibinfo {author} {\bibfnamefont {G.}~\bibnamefont
  {Alexandrowicz}}\ and\ \bibinfo {author} {\bibfnamefont {A.~P.}\ \bibnamefont
  {Jardine}},\ }\href {http://stacks.iop.org/0953-8984/19/i=30/a=305001}
  {\bibfield  {journal} {\bibinfo  {journal} {J. Phys.: Cond. Matt.}\ }\textbf
  {\bibinfo {volume} {19}},\ \bibinfo {pages} {305001} (\bibinfo {year}
  {2007})}\BibitemShut {NoStop}%
\bibitem [{\citenamefont {Jardine}\ \emph {et~al.}(2009)\citenamefont
  {Jardine}, \citenamefont {Hedgeland}, \citenamefont {Alexandrowicz},
  \citenamefont {Allison},\ and\ \citenamefont {Ellis}}]{Jardine2009}%
  \BibitemOpen
  \bibfield  {author} {\bibinfo {author} {\bibfnamefont {A.}~\bibnamefont
  {Jardine}}, \bibinfo {author} {\bibfnamefont {H.}~\bibnamefont {Hedgeland}},
  \bibinfo {author} {\bibfnamefont {G.}~\bibnamefont {Alexandrowicz}}, \bibinfo
  {author} {\bibfnamefont {W.}~\bibnamefont {Allison}}, \ and\ \bibinfo
  {author} {\bibfnamefont {J.}~\bibnamefont {Ellis}},\ }\href {\doibase
  {10.1016/j.progsurf.2009.07.001}} {\bibfield  {journal} {\bibinfo  {journal}
  {Prog. Surf. Sci.}\ }\textbf {\bibinfo {volume} {84}},\ \bibinfo {pages}
  {323} (\bibinfo {year} {2009})}\BibitemShut {NoStop}%
\bibitem [{\citenamefont {Hedgeland}\ \emph {et~al.}(2009)\citenamefont
  {Hedgeland}, \citenamefont {Fouquet}, \citenamefont {Jardine}, \citenamefont
  {Alexandrowicz}, \citenamefont {Allison},\ and\ \citenamefont
  {Ellis}}]{Hedgeland2009}%
  \BibitemOpen
  \bibfield  {author} {\bibinfo {author} {\bibfnamefont {H.}~\bibnamefont
  {Hedgeland}}, \bibinfo {author} {\bibfnamefont {P.}~\bibnamefont {Fouquet}},
  \bibinfo {author} {\bibfnamefont {A.~P.}\ \bibnamefont {Jardine}}, \bibinfo
  {author} {\bibfnamefont {G.}~\bibnamefont {Alexandrowicz}}, \bibinfo {author}
  {\bibfnamefont {W.}~\bibnamefont {Allison}}, \ and\ \bibinfo {author}
  {\bibfnamefont {J.}~\bibnamefont {Ellis}},\ }\href
  {http://dx.doi.org/10.1038/nphys1335} {\bibfield  {journal} {\bibinfo
  {journal} {Nat. Phys.}\ }\textbf {\bibinfo {volume} {5}},\ \bibinfo {pages}
  {561} (\bibinfo {year} {2009})}\BibitemShut {NoStop}%
\bibitem [{\citenamefont {Fouquet}, \citenamefont {Hedgeland},\ and\
  \citenamefont {Jardine}(2010)}]{Fouquet2010}%
  \BibitemOpen
  \bibfield  {author} {\bibinfo {author} {\bibfnamefont {P.}~\bibnamefont
  {Fouquet}}, \bibinfo {author} {\bibfnamefont {H.}~\bibnamefont {Hedgeland}},
  \ and\ \bibinfo {author} {\bibfnamefont {A.~P.}\ \bibnamefont {Jardine}},\
  }\href {\doibase 10.1524/zpch.2010.6092} {\bibfield  {journal} {\bibinfo
  {journal} {Z. Phys. Chem}\ }\textbf {\bibinfo {volume} {224}},\ \bibinfo
  {pages} {61} (\bibinfo {year} {2010})}\BibitemShut {NoStop}%
\bibitem [{\citenamefont {Tamt\"ogl}\ \emph {et~al.}(2018)\citenamefont
  {Tamt\"ogl}, \citenamefont {Sacchi}, \citenamefont {Calvo-Almazán},
  \citenamefont {Zbiri}, \citenamefont {Koza}, \citenamefont {Ernst},\ and\
  \citenamefont {Fouquet}}]{Tamtogl2018}%
  \BibitemOpen
  \bibfield  {author} {\bibinfo {author} {\bibfnamefont {A.}~\bibnamefont
  {Tamt\"ogl}}, \bibinfo {author} {\bibfnamefont {M.}~\bibnamefont {Sacchi}},
  \bibinfo {author} {\bibfnamefont {I.}~\bibnamefont {Calvo-Almazán}},
  \bibinfo {author} {\bibfnamefont {M.}~\bibnamefont {Zbiri}}, \bibinfo
  {author} {\bibfnamefont {M.~M.}\ \bibnamefont {Koza}}, \bibinfo {author}
  {\bibfnamefont {W.~E.}\ \bibnamefont {Ernst}}, \ and\ \bibinfo {author}
  {\bibfnamefont {P.}~\bibnamefont {Fouquet}},\ }\href {\doibase
  https://doi.org/10.1016/j.carbon.2017.09.104} {\bibfield  {journal} {\bibinfo
   {journal} {Carbon}\ }\textbf {\bibinfo {volume} {126}},\ \bibinfo {pages}
  {23} (\bibinfo {year} {2018})}\BibitemShut {NoStop}%
\bibitem [{\citenamefont {DeKieviet}\ \emph {et~al.}(1995)\citenamefont
  {DeKieviet}, \citenamefont {Dubbers}, \citenamefont {Schmidt}, \citenamefont
  {Scholz},\ and\ \citenamefont {Spinola}}]{DeKieviet1995}%
  \BibitemOpen
  \bibfield  {author} {\bibinfo {author} {\bibfnamefont {M.}~\bibnamefont
  {DeKieviet}}, \bibinfo {author} {\bibfnamefont {D.}~\bibnamefont {Dubbers}},
  \bibinfo {author} {\bibfnamefont {C.}~\bibnamefont {Schmidt}}, \bibinfo
  {author} {\bibfnamefont {D.}~\bibnamefont {Scholz}}, \ and\ \bibinfo {author}
  {\bibfnamefont {U.}~\bibnamefont {Spinola}},\ }\href {\doibase
  10.1103/PhysRevLett.75.1919} {\bibfield  {journal} {\bibinfo  {journal}
  {Phys. Rev. Lett.}\ }\textbf {\bibinfo {volume} {75}},\ \bibinfo {pages}
  {1919} (\bibinfo {year} {1995})}\BibitemShut {NoStop}%
\bibitem [{\citenamefont {Mezei}, \citenamefont {Pappas},\ and\ \citenamefont
  {Gutberlet}(2003)}]{Mezei2003}%
  \BibitemOpen
  \bibfield  {author} {\bibinfo {author} {\bibfnamefont {F.}~\bibnamefont
  {Mezei}}, \bibinfo {author} {\bibfnamefont {C.}~\bibnamefont {Pappas}}, \
  and\ \bibinfo {author} {\bibfnamefont {T.}~\bibnamefont {Gutberlet}},\ }\href
  {\doibase https://doi.org/10.1007/3-540-45823-9} {\emph {\bibinfo {title}
  {{Neutron spin echo spectroscopy: basics, trends, and applications}}}},\
  \bibinfo {series} {Lecture notes in physics}\ No.\ \bibinfo {number} {601}\
  (\bibinfo  {publisher} {Springer},\ \bibinfo {address} {Berlin ; New York},\
  \bibinfo {year} {2003})\BibitemShut {NoStop}%
\bibitem [{\citenamefont {Far\'{i}as}\ and\ \citenamefont
  {Rieder}(1998)}]{Farias1998}%
  \BibitemOpen
  \bibfield  {author} {\bibinfo {author} {\bibfnamefont {D.}~\bibnamefont
  {Far\'{i}as}}\ and\ \bibinfo {author} {\bibfnamefont {K.-H.}\ \bibnamefont
  {Rieder}},\ }\href {http://stacks.iop.org/0034-4885/61/i=12/a=001} {\bibfield
   {journal} {\bibinfo  {journal} {Rep. Prog. Phys.}\ }\textbf {\bibinfo
  {volume} {61}},\ \bibinfo {pages} {1575} (\bibinfo {year}
  {1998})}\BibitemShut {NoStop}%
\bibitem [{\citenamefont {Benedek}\ and\ \citenamefont
  {Toennies}(1994)}]{Benedek1994}%
  \BibitemOpen
  \bibfield  {author} {\bibinfo {author} {\bibfnamefont {G.}~\bibnamefont
  {Benedek}}\ and\ \bibinfo {author} {\bibfnamefont {J.~P.}\ \bibnamefont
  {Toennies}},\ }\href@noop {} {\bibfield  {journal} {\bibinfo  {journal}
  {Surf. Sci.}\ }\textbf {\bibinfo {volume} {299}},\ \bibinfo {pages} {587}
  (\bibinfo {year} {1994})}\BibitemShut {NoStop}%
\bibitem [{\citenamefont {Kraus}\ \emph {et~al.}(2013)\citenamefont {Kraus},
  \citenamefont {Tamt\"ogl}, \citenamefont {Mayrhofer-Reinhartshuber},
  \citenamefont {Benedek},\ and\ \citenamefont {Ernst}}]{Kraus2013}%
  \BibitemOpen
  \bibfield  {author} {\bibinfo {author} {\bibfnamefont {P.}~\bibnamefont
  {Kraus}}, \bibinfo {author} {\bibfnamefont {A.}~\bibnamefont {Tamt\"ogl}},
  \bibinfo {author} {\bibfnamefont {M.}~\bibnamefont
  {Mayrhofer-Reinhartshuber}}, \bibinfo {author} {\bibfnamefont
  {G.}~\bibnamefont {Benedek}}, \ and\ \bibinfo {author} {\bibfnamefont
  {W.~E.}\ \bibnamefont {Ernst}},\ }\href {\doibase 10.1103/PhysRevB.87.245433}
  {\bibfield  {journal} {\bibinfo  {journal} {Phys. Rev. B}\ }\textbf {\bibinfo
  {volume} {87}},\ \bibinfo {pages} {245433} (\bibinfo {year}
  {2013})}\BibitemShut {NoStop}%
\bibitem [{\citenamefont {G\"ahler}\ \emph {et~al.}(1996)\citenamefont
  {G\"ahler}, \citenamefont {Golub}, \citenamefont {Habicht}, \citenamefont
  {Keller},\ and\ \citenamefont {Felber}}]{Gaehler1996}%
  \BibitemOpen
  \bibfield  {author} {\bibinfo {author} {\bibfnamefont {R.}~\bibnamefont
  {G\"ahler}}, \bibinfo {author} {\bibfnamefont {R.}~\bibnamefont {Golub}},
  \bibinfo {author} {\bibfnamefont {K.}~\bibnamefont {Habicht}}, \bibinfo
  {author} {\bibfnamefont {T.}~\bibnamefont {Keller}}, \ and\ \bibinfo {author}
  {\bibfnamefont {J.}~\bibnamefont {Felber}},\ }\href {\doibase
  https://doi.org/10.1016/S0921-4526(96)00509-1} {\bibfield  {journal}
  {\bibinfo  {journal} {Physica B Condens Matter.}\ }\textbf {\bibinfo {volume}
  {229}},\ \bibinfo {pages} {1} (\bibinfo {year} {1996})}\BibitemShut {NoStop}%
\bibitem [{\citenamefont {Bahn}\ \emph {et~al.}(2017)\citenamefont {Bahn},
  \citenamefont {Tamt\"{o}gl}, \citenamefont {Ellis}, \citenamefont {Allison},\
  and\ \citenamefont {Fouquet}}]{Bahn2017}%
  \BibitemOpen
  \bibfield  {author} {\bibinfo {author} {\bibfnamefont {E.}~\bibnamefont
  {Bahn}}, \bibinfo {author} {\bibfnamefont {A.}~\bibnamefont {Tamt\"{o}gl}},
  \bibinfo {author} {\bibfnamefont {J.}~\bibnamefont {Ellis}}, \bibinfo
  {author} {\bibfnamefont {W.}~\bibnamefont {Allison}}, \ and\ \bibinfo
  {author} {\bibfnamefont {P.}~\bibnamefont {Fouquet}},\ }\href {\doibase
  10.1016/j.carbon.2016.12.055} {\bibfield  {journal} {\bibinfo  {journal}
  {Carbon}\ }\textbf {\bibinfo {volume} {114}},\ \bibinfo {pages} {504}
  (\bibinfo {year} {2017})}\BibitemShut {NoStop}%
\bibitem [{\citenamefont {Tamt\"ogl}\ \emph {et~al.}(2015)\citenamefont
  {Tamt\"ogl}, \citenamefont {Bahn}, \citenamefont {Zhu}, \citenamefont
  {Fouquet}, \citenamefont {Ellis},\ and\ \citenamefont
  {Allison}}]{Tamtogl2015}%
  \BibitemOpen
  \bibfield  {author} {\bibinfo {author} {\bibfnamefont {A.}~\bibnamefont
  {Tamt\"ogl}}, \bibinfo {author} {\bibfnamefont {E.}~\bibnamefont {Bahn}},
  \bibinfo {author} {\bibfnamefont {J.}~\bibnamefont {Zhu}}, \bibinfo {author}
  {\bibfnamefont {P.}~\bibnamefont {Fouquet}}, \bibinfo {author} {\bibfnamefont
  {J.}~\bibnamefont {Ellis}}, \ and\ \bibinfo {author} {\bibfnamefont
  {W.}~\bibnamefont {Allison}},\ }\href
  {http://dx.doi.org/10.1021/acs.jpcc.5b08284} {\bibfield  {journal} {\bibinfo
  {journal} {J. Phys. Chem. C}\ }\textbf {\bibinfo {volume} {119}},\ \bibinfo
  {pages} {25983} (\bibinfo {year} {2015})}\BibitemShut {NoStop}%
\bibitem [{\citenamefont {Tamt\"ogl}\ \emph {et~al.}(2017)\citenamefont
  {Tamt\"ogl}, \citenamefont {Kraus}, \citenamefont {Avidor}, \citenamefont
  {Bremholm}, \citenamefont {Hedegaard}, \citenamefont {Iversen}, \citenamefont
  {Bianchi}, \citenamefont {Hofmann}, \citenamefont {Ellis}, \citenamefont
  {Allison}, \citenamefont {Benedek},\ and\ \citenamefont
  {Ernst}}]{Tamtogl2017}%
  \BibitemOpen
  \bibfield  {author} {\bibinfo {author} {\bibfnamefont {A.}~\bibnamefont
  {Tamt\"ogl}}, \bibinfo {author} {\bibfnamefont {P.}~\bibnamefont {Kraus}},
  \bibinfo {author} {\bibfnamefont {N.}~\bibnamefont {Avidor}}, \bibinfo
  {author} {\bibfnamefont {M.}~\bibnamefont {Bremholm}}, \bibinfo {author}
  {\bibfnamefont {E.~M.~J.}\ \bibnamefont {Hedegaard}}, \bibinfo {author}
  {\bibfnamefont {B.~B.}\ \bibnamefont {Iversen}}, \bibinfo {author}
  {\bibfnamefont {M.}~\bibnamefont {Bianchi}}, \bibinfo {author} {\bibfnamefont
  {P.}~\bibnamefont {Hofmann}}, \bibinfo {author} {\bibfnamefont
  {J.}~\bibnamefont {Ellis}}, \bibinfo {author} {\bibfnamefont
  {W.}~\bibnamefont {Allison}}, \bibinfo {author} {\bibfnamefont
  {G.}~\bibnamefont {Benedek}}, \ and\ \bibinfo {author} {\bibfnamefont
  {W.~E.}\ \bibnamefont {Ernst}},\ }\href {\doibase 10.1103/PhysRevB.95.195401}
  {\bibfield  {journal} {\bibinfo  {journal} {Phys. Rev. B}\ }\textbf {\bibinfo
  {volume} {95}},\ \bibinfo {pages} {195401} (\bibinfo {year}
  {2017})}\BibitemShut {NoStop}%
\bibitem [{\citenamefont {Jardine}\ \emph {et~al.}(2004)\citenamefont
  {Jardine}, \citenamefont {Dworski}, \citenamefont {Fouquet}, \citenamefont
  {Alexandrowicz}, \citenamefont {Riley}, \citenamefont {Lee}, \citenamefont
  {Ellis},\ and\ \citenamefont {Allison}}]{Jardine2004}%
  \BibitemOpen
  \bibfield  {author} {\bibinfo {author} {\bibfnamefont {A.~P.}\ \bibnamefont
  {Jardine}}, \bibinfo {author} {\bibfnamefont {S.}~\bibnamefont {Dworski}},
  \bibinfo {author} {\bibfnamefont {P.}~\bibnamefont {Fouquet}}, \bibinfo
  {author} {\bibfnamefont {G.}~\bibnamefont {Alexandrowicz}}, \bibinfo {author}
  {\bibfnamefont {D.~J.}\ \bibnamefont {Riley}}, \bibinfo {author}
  {\bibfnamefont {G.~Y.~H.}\ \bibnamefont {Lee}}, \bibinfo {author}
  {\bibfnamefont {J.}~\bibnamefont {Ellis}}, \ and\ \bibinfo {author}
  {\bibfnamefont {W.}~\bibnamefont {Allison}},\ }\href
  {http://www.sciencemag.org/content/304/5678/1790.abstract} {\bibfield
  {journal} {\bibinfo  {journal} {Science}\ }\textbf {\bibinfo {volume}
  {304}},\ \bibinfo {pages} {1790} (\bibinfo {year} {2004})}\BibitemShut
  {NoStop}%
\bibitem [{\citenamefont {Fouquet}\ \emph {et~al.}(2005)\citenamefont
  {Fouquet}, \citenamefont {Jardine}, \citenamefont {Dworski}, \citenamefont
  {Alexandrowicz}, \citenamefont {Allison},\ and\ \citenamefont
  {Ellis}}]{Fouquet2005}%
  \BibitemOpen
  \bibfield  {author} {\bibinfo {author} {\bibfnamefont {P.}~\bibnamefont
  {Fouquet}}, \bibinfo {author} {\bibfnamefont {A.~P.}\ \bibnamefont
  {Jardine}}, \bibinfo {author} {\bibfnamefont {S.}~\bibnamefont {Dworski}},
  \bibinfo {author} {\bibfnamefont {G.}~\bibnamefont {Alexandrowicz}}, \bibinfo
  {author} {\bibfnamefont {W.}~\bibnamefont {Allison}}, \ and\ \bibinfo
  {author} {\bibfnamefont {J.}~\bibnamefont {Ellis}},\ }\href {\doibase
  10.1063/1.1896945} {\bibfield  {journal} {\bibinfo  {journal} {Rev. Sci.
  Instrum.}\ }\textbf {\bibinfo {volume} {76}},\ \bibinfo {pages} {053109}
  (\bibinfo {year} {2005})}\BibitemShut {NoStop}%
\bibitem [{\citenamefont {Farago}(2003)}]{Farago2003}%
  \BibitemOpen
  \bibfield  {author} {\bibinfo {author} {\bibfnamefont {B.}~\bibnamefont
  {Farago}},\ }\enquote {\bibinfo {title} {Ill neutron data booklet},}\ \
  (\bibinfo  {publisher} {OCP Science},\ \bibinfo {year} {2003})\ Chap.\
  \bibinfo {chapter} {The basics of Neutron Spin Echo}, pp.\ \bibinfo {pages}
  {91--106}\BibitemShut {NoStop}%
\bibitem [{\citenamefont {Ohl}\ \emph {et~al.}(2012)\citenamefont {Ohl},
  \citenamefont {Monkenbusch}, \citenamefont {Arend}, \citenamefont
  {Kozielewski}, \citenamefont {Vehres}, \citenamefont {Tiemann}, \citenamefont
  {Butzek}, \citenamefont {Soltner}, \citenamefont {Giesen}, \citenamefont
  {Achten}, \citenamefont {Stelzer}, \citenamefont {Lindenau}, \citenamefont
  {Budwig}, \citenamefont {Kleines}, \citenamefont {Drochner}, \citenamefont
  {Kaemmerling}, \citenamefont {Wagener}, \citenamefont {M\"oller},
  \citenamefont {Iverson}, \citenamefont {Sharp},\ and\ \citenamefont
  {Richter}}]{Ohl2012}%
  \BibitemOpen
  \bibfield  {author} {\bibinfo {author} {\bibfnamefont {M.}~\bibnamefont
  {Ohl}}, \bibinfo {author} {\bibfnamefont {M.}~\bibnamefont {Monkenbusch}},
  \bibinfo {author} {\bibfnamefont {N.}~\bibnamefont {Arend}}, \bibinfo
  {author} {\bibfnamefont {T.}~\bibnamefont {Kozielewski}}, \bibinfo {author}
  {\bibfnamefont {G.}~\bibnamefont {Vehres}}, \bibinfo {author} {\bibfnamefont
  {C.}~\bibnamefont {Tiemann}}, \bibinfo {author} {\bibfnamefont
  {M.}~\bibnamefont {Butzek}}, \bibinfo {author} {\bibfnamefont
  {H.}~\bibnamefont {Soltner}}, \bibinfo {author} {\bibfnamefont
  {U.}~\bibnamefont {Giesen}}, \bibinfo {author} {\bibfnamefont
  {R.}~\bibnamefont {Achten}}, \bibinfo {author} {\bibfnamefont
  {H.}~\bibnamefont {Stelzer}}, \bibinfo {author} {\bibfnamefont
  {B.}~\bibnamefont {Lindenau}}, \bibinfo {author} {\bibfnamefont
  {A.}~\bibnamefont {Budwig}}, \bibinfo {author} {\bibfnamefont
  {H.}~\bibnamefont {Kleines}}, \bibinfo {author} {\bibfnamefont
  {M.}~\bibnamefont {Drochner}}, \bibinfo {author} {\bibfnamefont
  {P.}~\bibnamefont {Kaemmerling}}, \bibinfo {author} {\bibfnamefont
  {M.}~\bibnamefont {Wagener}}, \bibinfo {author} {\bibfnamefont
  {R.}~\bibnamefont {M\"oller}}, \bibinfo {author} {\bibfnamefont
  {E.}~\bibnamefont {Iverson}}, \bibinfo {author} {\bibfnamefont
  {M.}~\bibnamefont {Sharp}}, \ and\ \bibinfo {author} {\bibfnamefont
  {D.}~\bibnamefont {Richter}},\ }\href {\doibase
  http://dx.doi.org/10.1016/j.nima.2012.08.059} {\bibfield  {journal} {\bibinfo
   {journal} {Nucl. Instr. Meth. Phys. Res.}\ }\textbf {\bibinfo {volume}
  {696}},\ \bibinfo {pages} {85 } (\bibinfo {year} {2012})}\BibitemShut
  {NoStop}%
\bibitem [{\citenamefont {Jones}\ \emph {et~al.}(2016)\citenamefont {Jones},
  \citenamefont {Tamt\"ogl}, \citenamefont {Calvo-Almaz\'an},\ and\
  \citenamefont {Hansen}}]{Jones2016}%
  \BibitemOpen
  \bibfield  {author} {\bibinfo {author} {\bibfnamefont {A.}~\bibnamefont
  {Jones}}, \bibinfo {author} {\bibfnamefont {A.}~\bibnamefont {Tamt\"ogl}},
  \bibinfo {author} {\bibfnamefont {I.}~\bibnamefont {Calvo-Almaz\'an}}, \ and\
  \bibinfo {author} {\bibfnamefont {A.}~\bibnamefont {Hansen}},\ }\href
  {\doibase http://dx.doi.org/10.1038/srep27776} {\bibfield  {journal}
  {\bibinfo  {journal} {Sci. Rep.}\ }\textbf {\bibinfo {volume} {6}},\ \bibinfo
  {pages} {27776} (\bibinfo {year} {2016})}\BibitemShut {NoStop}%
\bibitem [{\citenamefont {Ward}(2013)}]{Ward2013}%
  \BibitemOpen
  \bibfield  {author} {\bibinfo {author} {\bibfnamefont {D.~J.}\ \bibnamefont
  {Ward}},\ }\emph {\bibinfo {title} {{A Study of Spin/echo Lineshapes in
  Helium Atom Scattering from Adsorbates}}},\ \href@noop {} {Ph.D. thesis},\
  \bibinfo  {school} {University of Cambridge} (\bibinfo {year}
  {2013})\BibitemShut {NoStop}%
\bibitem [{\citenamefont {Caldwell}(1977)}]{Caldwell1977}%
  \BibitemOpen
  \bibfield  {author} {\bibinfo {author} {\bibfnamefont {C.~D.}\ \bibnamefont
  {Caldwell}},\ }\href {\doibase 10.1364/OL.1.000101} {\bibfield  {journal}
  {\bibinfo  {journal} {Opt. Lett.}\ }\textbf {\bibinfo {volume} {1}},\
  \bibinfo {pages} {101} (\bibinfo {year} {1977})}\BibitemShut {NoStop}%
\bibitem [{\citenamefont {Pippard}(1985)}]{Pippard1985}%
  \BibitemOpen
  \bibfield  {author} {\bibinfo {author} {\bibfnamefont {A.~B.}\ \bibnamefont
  {Pippard}},\ }\href@noop {} {\emph {\bibinfo {title} {{Response and
  Stability}}}}\ (\bibinfo  {publisher} {Cambridge University Press, UK},\
  \bibinfo {year} {1985})\BibitemShut {NoStop}%
\bibitem [{\citenamefont {Taylor}(1994)}]{Taylor1994}%
  \BibitemOpen
  \bibfield  {author} {\bibinfo {author} {\bibfnamefont {F.~J.}\ \bibnamefont
  {Taylor}},\ }\href@noop {} {\emph {\bibinfo {title} {{Principals of Signals
  and Systems}}}},\ edited by\ \bibinfo {editor} {\bibfnamefont
  {H.}~\bibnamefont {G.T.}}\ and\ \bibinfo {editor} {\bibfnamefont {M.~J.}\
  \bibnamefont {M.}}\ (\bibinfo  {publisher} {McGraw Hill},\ \bibinfo {year}
  {1994})\BibitemShut {NoStop}%
\bibitem [{\citenamefont {Gonz\`{a}lez}\ \emph {et~al.}(2007)\citenamefont
  {Gonz\`{a}lez}, \citenamefont {Santiago}, \citenamefont {Slezak},\ and\
  \citenamefont {Peuriot}}]{Gonzalez2007}%
  \BibitemOpen
  \bibfield  {author} {\bibinfo {author} {\bibfnamefont {M.~G.}\ \bibnamefont
  {Gonz\`{a}lez}}, \bibinfo {author} {\bibfnamefont {G.~D.}\ \bibnamefont
  {Santiago}}, \bibinfo {author} {\bibfnamefont {V.~B.}\ \bibnamefont
  {Slezak}}, \ and\ \bibinfo {author} {\bibfnamefont {A.~L.}\ \bibnamefont
  {Peuriot}},\ }\href {\doibase http://dx.doi.org/10.1063/1.2740063} {\bibfield
   {journal} {\bibinfo  {journal} {Rev. Sci. Instr.}\ }\textbf {\bibinfo
  {volume} {78}},\ \bibinfo {pages} {055108} (\bibinfo {year}
  {2007})}\BibitemShut {NoStop}%
\bibitem [{\citenamefont {Schultz}(2016)}]{Schultz2016}%
  \BibitemOpen
  \bibfield  {author} {\bibinfo {author} {\bibfnamefont {K.~D.}\ \bibnamefont
  {Schultz}},\ }\href {\doibase 10.1119/1.4953341} {\bibfield  {journal}
  {\bibinfo  {journal} {Am. J. Phys}\ }\textbf {\bibinfo {volume} {84}},\
  \bibinfo {pages} {557} (\bibinfo {year} {2016})}\BibitemShut {NoStop}%
\bibitem [{\citenamefont {McIntosh}\ \emph {et~al.}(2018)\citenamefont
  {McIntosh}, \citenamefont {Tamt\"ogl}, \citenamefont {Ward}, \citenamefont
  {Hedgeland}, \citenamefont {Jardine}, \citenamefont {Ellis},\ and\
  \citenamefont {Allison}}]{McIntosh2016}%
  \BibitemOpen
  \bibfield  {author} {\bibinfo {author} {\bibfnamefont {E.~M.}\ \bibnamefont
  {McIntosh}}, \bibinfo {author} {\bibfnamefont {A.}~\bibnamefont {Tamt\"ogl}},
  \bibinfo {author} {\bibfnamefont {D.~J.}\ \bibnamefont {Ward}}, \bibinfo
  {author} {\bibfnamefont {H.}~\bibnamefont {Hedgeland}}, \bibinfo {author}
  {\bibfnamefont {A.~P.}\ \bibnamefont {Jardine}}, \bibinfo {author}
  {\bibfnamefont {J.}~\bibnamefont {Ellis}}, \ and\ \bibinfo {author}
  {\bibfnamefont {W.}~\bibnamefont {Allison}},\ }\href@noop {} {\  (\bibinfo
  {year} {2018})},\ \bibinfo {note} {unpublished}\BibitemShut {NoStop}%
\bibitem [{\citenamefont {Lechner}\ \emph {et~al.}(2013)\citenamefont
  {Lechner}, \citenamefont {Hedgeland}, \citenamefont {Allison}, \citenamefont
  {Ellis},\ and\ \citenamefont {Jardine}}]{Lechner2013}%
  \BibitemOpen
  \bibfield  {author} {\bibinfo {author} {\bibfnamefont {B.~A.~J.}\
  \bibnamefont {Lechner}}, \bibinfo {author} {\bibfnamefont {H.}~\bibnamefont
  {Hedgeland}}, \bibinfo {author} {\bibfnamefont {W.}~\bibnamefont {Allison}},
  \bibinfo {author} {\bibfnamefont {J.}~\bibnamefont {Ellis}}, \ and\ \bibinfo
  {author} {\bibfnamefont {A.~P.}\ \bibnamefont {Jardine}},\ }\href {\doibase
  http://dx.doi.org/10.1063/1.4791929} {\bibfield  {journal} {\bibinfo
  {journal} {Rev. Sci. Instr.}\ }\textbf {\bibinfo {volume} {84}},\ \bibinfo
  {pages} {026105} (\bibinfo {year} {2013})}\BibitemShut {NoStop}%
\end{thebibliography}%

\clearpage
\onecolumngrid

\renewcommand{\thepage}{\arabic{apppage}}
\pagenumbering{arabic}

\setcounter{section}{0}
\renewcommand{\thesection}{S\arabic{section}}
\enlargethispage{4\baselineskip}
\subsection*{\large{Polarisation in Spin-Echo Experiments: Multi-point and Lock-in Measurements:\\ Supplementary Information}}

\setcounter{equation}{0}
\setcounter{figure}{0}
\makeatletter
\renewcommand{\theequation}{S\arabic{equation}}
\renewcommand{\thefigure}{S\arabic{figure}}

\section{Measurement of the polarised signal in the $^3$He beam}
In an ideal experiment the accumulated phase of a $^3$He atom of wavelength $\lambda$, travelling parallel to a field, $B$ along $z$ is given in Eq. [1] of \footnote{\label{note1}G. Alexandrowicz and A. P. Jardine, J. Phys.: Cond. Matt. \textbf{19}, 305001 (2007).} as
\begin{equation}
\phi = \frac{\gamma m \lambda}{h} \int_0^L B(z) \, \mathrm{d} z = \frac{\gamma m \lambda}{h} \underbrace{B_{eff}}_{\int_0^L \frac{ B(z) }{I} \, \mathrm{d} z } I  = \kappa I \lambda \; ,
\label{eq:phase}
\end{equation}
where $\gamma$ is the gyromagnetic ratio and $m$ is the particle mass. The field integral along the path through the phase-coil corresponds to an effective field per unit energising current $B_{eff}$, hence the constant $\kappa = \gamma m B_{eff} / h$. As the beam passes through the phase coil, the overall polarisation is an integral over the wavelength distribution in the beam and is given by (Eq. [2] of \textsuperscript{\ref{note1}})
\begin{equation}
P = P_x (I) + \mathbbm{i} P_y (I) = \int \rho(\lambda) \, e^{2 \pi \mathbbm{i} \kappa I \lambda } \, \mathrm{d} \lambda \; ,
\label{eq:Fourier}
\end{equation}
where $\rho(\lambda)$ defines the wavelength distribution. The wavelength distribution in the incident beam may be approximated by a Gaussian function with a mean wavelength, $\lambda_0$ and width, $\sigma_{\lambda}$, so that
\begin{equation} 
\rho(\lambda) = \tfrac{1}{\sqrt{2 \pi} \sigma_{\lambda} } \, \exp \left( - \tfrac{ \left( \lambda - \lambda_0 \right)^2 }{ 2 \sigma_{\lambda}^2 } \right) \, .
\label{eq:rho}
\end{equation}
Inserting \eqref{eq:rho} in \eqref{eq:Fourier} and performing the Fourier transform gives
\begin{equation} 
\begin{aligned}
P (I) &= \tfrac{1}{\sqrt{2 \pi} \sigma_{\lambda} } \, \int \exp \left( - \tfrac{ \left( \lambda - \lambda_0 \right)^2 }{ 2 \sigma_{\lambda}^2 } \right) \, \exp \left( 2 \pi \mathbbm{i} \kappa I \lambda \right) \, \mathrm{d} \lambda \\
 &= \exp \left( 2 \pi^2 \sigma_{\lambda}^2 \kappa^2 I^2 \right) \; \exp \left( 2 \pi \mathbbm{i} \kappa I \lambda_0 \right) \,.
\label{eq:Fourier2}
\end{aligned}
\end{equation}
And the real part of the polarisation in this idealised experiment is
\begin{equation}
\begin{aligned}
P_x (I) &=  \exp \left( - 2 \pi^2 \sigma_{\lambda}^2 \kappa^2 I^2 \right) \, \cos \left( 2 \pi \kappa I \lambda_0 \right) \\
 &=  \exp \left( - \tfrac{ I^2 } {2 \sigma^2 } \right) \, \cos \left( \Omega_0 I \right) \; ,
\label{eq:Pol}
\end{aligned}
\end{equation}
where $\Omega_0 = 2 \pi \kappa \lambda_0$ and $\sigma= 1/(2 \pi \sigma_{\lambda} \kappa)$. In the case of a real experiment, \eqref{eq:Pol} is modified in three ways. First, the signal passed by the polariser is proportional to the detector sensitivity, which introduces a multiplicative constant, $A$. Second, the possibility of mechanical misalignment and stray magnetic fields may introduce a phase factor, $\delta$. Finally there is a static offset, $C$, arising from the unpolarised component in the beam together with any background signal in the detector. Hence we obtain the form described in the main text
\begin{equation}
P_{x}(I) = A \; \exp \left( - \tfrac{ I^2 } {2 \sigma^2 } \right) \, \cos \left[ \Omega_0 (I - \delta) \right] +  C \, .
\label{equ:phasecoil2}
\end{equation}

\section{Circuit Diagrams}
\subsection{Output Circuit}
\begin{figure*}[htb]
\centering
 \includegraphics[width=0.95\textwidth]{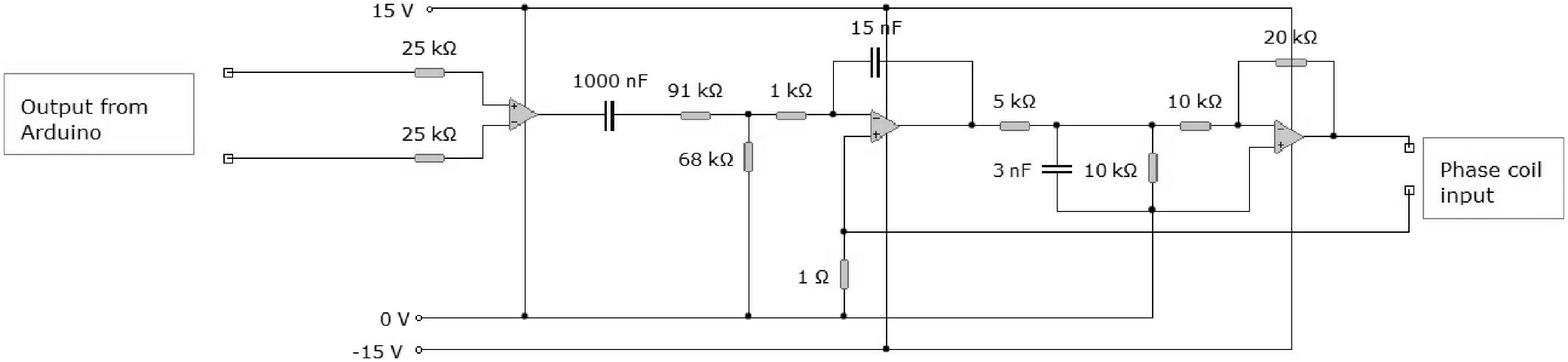}
 \caption{Circuit diagram for converting the output voltage from the DAC to the phase coil current. An initial buffer using an INA105 chip ensures only small currents are drawn from the Arduino. The $1000~$nF capacitor removes the DC offset. It is followed by a potential divider and the remaining components form a compound, current-boosting amplifier to convert the voltage to a current in the phase-coil. The op-amps are, respectively AD8675 and OPA544, the latter is mounted on a heat-sink. Conversion $I_{0ut} / V_{Arduino} =0.428~\mathrm{A}/\mathrm{V}$.\label{fig:outputcircuit}}
\end{figure*}

The output circuit was required to change a $0-5$ V input to a $-1$ A to $+1$ A output current. Initially a buffer was used to ensure no current was drawn from the Arduino. The signal then passed through an RC filter to remove the D.C. offset. A potential divider was used to scale the voltage to the correct range followed by a series of amplifiers to scale the voltage to the right current and ensure the current was being drawn from an external power supply (\autoref{fig:outputcircuit}).

\subsection{Input Circuit}
\begin{figure*}[htb]
\centering
 \includegraphics[width=0.75\textwidth]{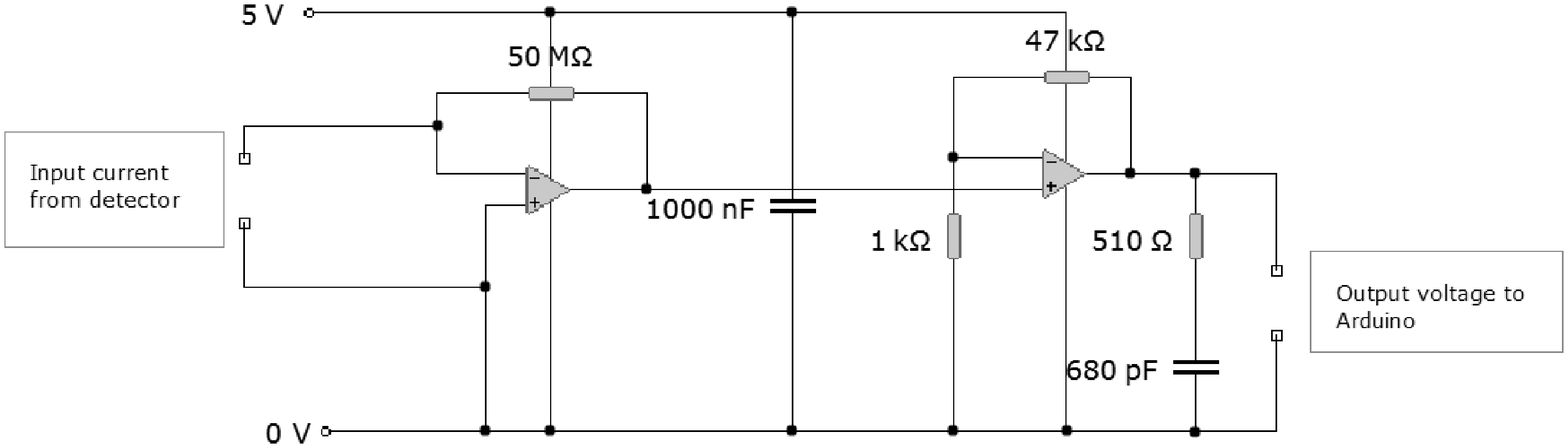}
 \caption{Circuit diagram for conversion of the input current from the detector into a voltage for the Arduino. The first amplifier is connected as an electrometer with gain $\Delta V_{out} / \Delta I_{In} = 5 \cdot 10^7$. The second amplifier buffers the electrometer and provides additional gain. The overall sensitivity is approximately $2.4 \cdot 10^9 ~\mathrm{V}/\mathrm{A}$.  Both amplifiers are precision, rail-to rail devices with a low input bias current ($< 1~\mathrm{pA}$) in a single package (AD8607).\label{fig:inputcircuit}}
\end{figure*}

The input current from the detector was between 0 and 1 nA and needed to be converted to a voltage between 0 and 5 V. The circuit designed for this (\autoref{fig:inputcircuit}) uses an amplifier to convert the incoming current into a voltage followed by another amplifier to provide the correct magnitude. As the signal is only a very small current, it is susceptible to background noise. The cable from the detector was soldered directly onto the leg of the first amplifier to reduce the distance over which noise could be picked up and the whole circuit was placed as close as possible to the detector output.\\
The fixed gain of the circuit in \autoref{fig:inputcircuit} and the limited resolution of the ADC in the Arduino limits the range over which the circuit is useful. It is sufficient to demonstrate the effectiveness of the method but would require a more sophisticated circuit to cover a dynamic range of 4 orders of magnitude observed in a typical experiment.

\newpage
\section{Numerical Control System and Models in Simulink}
\label{sec:Simulink}
The Simulink models used for the numerical models and hardware implementation are given in the figures below. \autoref{fig:modelfirstlevel} shows the polarisation control system used in both the numerical models and the real life implementation. Included within this system are the lock-in amplifier subsystem shown in \autoref{fig:lockinamp} and the polarisation calculator shown in \autoref{fig:polarisationcalculator}. In the numerical model the detector current input came from the output of the machine model given in \autoref{fig:machinemodel} and the phase coil current output is used as the input to the machine model.\\
\begin{figure}[htb]
\centering
\includegraphics[width=0.98\textwidth]{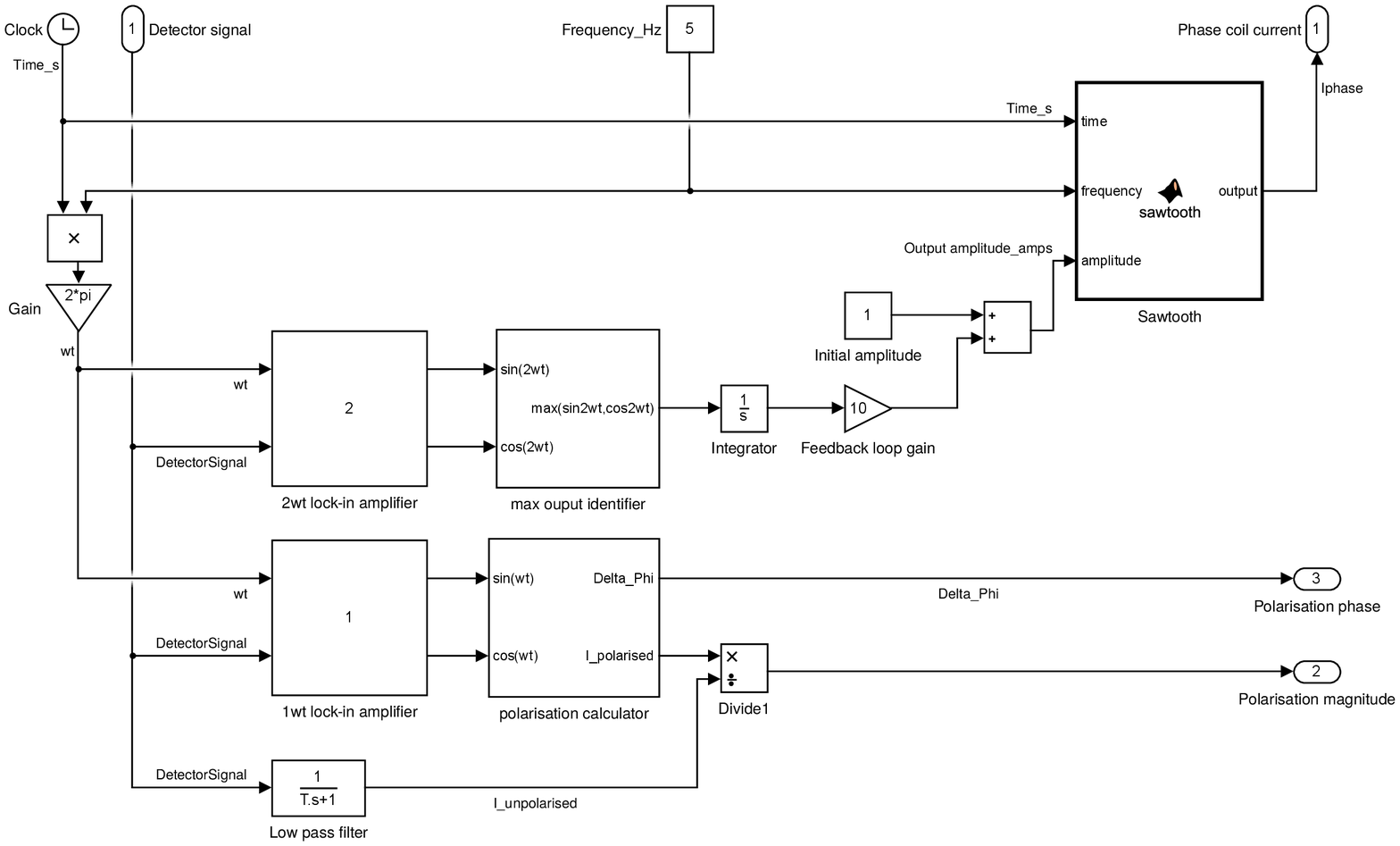}
\caption{The polarisation control system used. The model is based on the layout of \autoref{fig:blockdiagram} in the main part of the article. The detector signal arrives in the top left. It then enters the two lock-in amplifiers (described in \autoref{fig:lockinamp}) and a low pass filter. The $2\omega$ signal enters a new component which outputs only the maximum of either the sin or cos lock-in amplifier. This is then integrated and multiplied by the feedback loop gain. A Matlab script then creates a sawtooth wave with this amplitude with the output shown in the top right. The $1\omega$ signal and the unpolarised signal pass through the polarisation calculator which are illustrated in detail in \autoref{fig:polarisationcalculator}.}
\label{fig:modelfirstlevel}
\end{figure}
\begin{figure}[ht]
\centering
\includegraphics[width=0.8\textwidth]{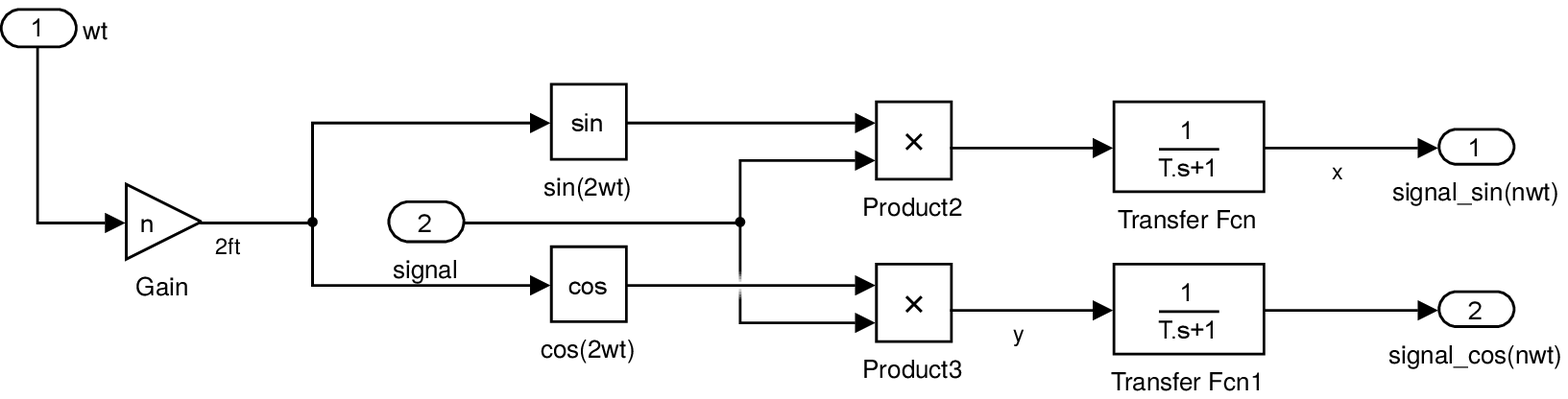}
\caption{The lock-in amplifier subsystem. The signal is multiplied separately by $\sin (n\omega t)$ and $\cos (n\omega t)$ then passed through a low pass filter.\label{fig:lockinamp}}
\end{figure}
The polarisation calculator (\autoref{fig:polarisationcalculator}) uses the dependence of the detector count rate $n_{det}$ upon the phase coil current $I$ according to \autoref{equ:phasecoil} as introduced in the main text. \eqref{equ:phasecoil} can be expanded to highlight the separation of the real and imaginary parts of the polarisation
\begin{equation}
n_{det}(I) = \mathrm{e} ^{(- I^2 / 2 \sigma^2 )} \left[ A_1 \cos(\Omega_0 I) + A_2 \sin(\Omega_0 I) \right] + C \; ,
\label{equ:phasecoil2}
\end{equation}
where $A_1$ and $A_2$ are the amplitudes of the real and imaginary part of the polarisation, respectively. The magnitude of the polarisation $P$ and the phase $\delta$ are then obtained via
\begin{equation}
|P| = \frac{\sqrt{A_1^2+A_2^2}}{C}\quad,\qquad  \delta = \tan^{-1}\left(\frac{A_2 }{A_1 }\right).
\label{equ:polarisation}
\end{equation}
\begin{figure}[ht]
\centering
\includegraphics[width=0.65\textwidth]{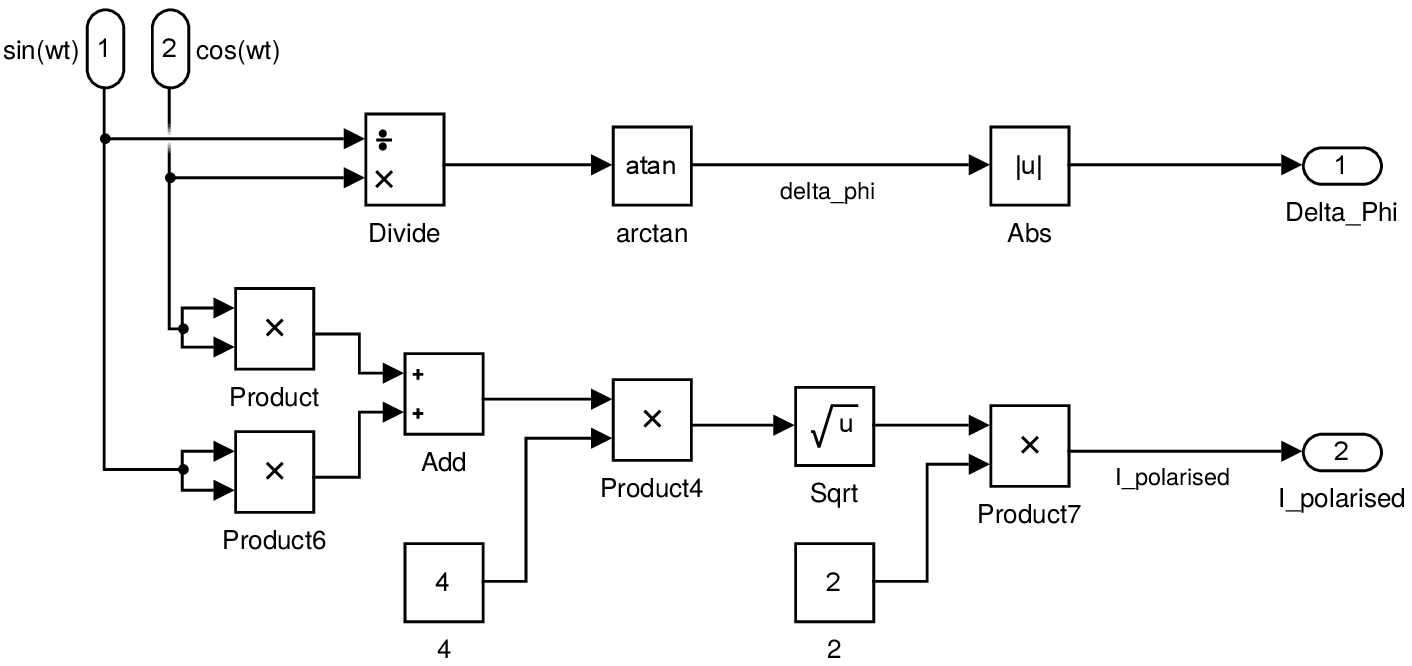}
\caption{Subsystem for calculating the polarisation and phase of the signal using \autoref{equ:polarisation}.\label{fig:polarisationcalculator}}
\end{figure}
\begin{figure}[ht]
\centering
\includegraphics[width=0.95\textwidth]{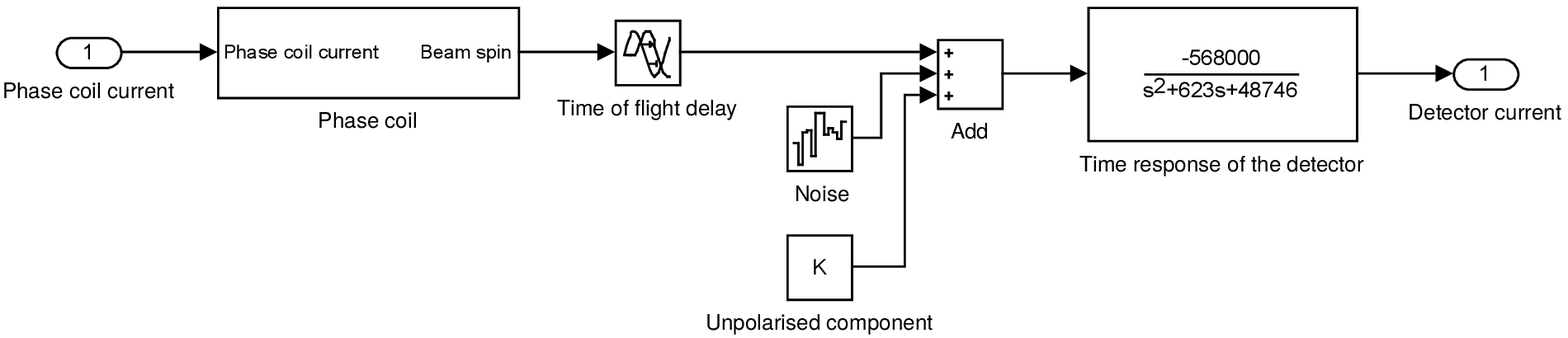}
\caption{The model used for the spin-echo machine. The phase coil current enters from the left. The phase coil subsystem shown in \autoref{fig:phasecoil} converts this to the number of $^3$He atoms arriving at the detector. It then passes through a delay modelling the time-of-flight, noise is added and the transfer function of the detector given by \autoref{equ:transferfunction} in the main text is applied. The output is then the current from the detector.}
\label{fig:machinemodel}
\end{figure}
\begin{figure}[ht]
\centering
\includegraphics[width=0.85\textwidth]{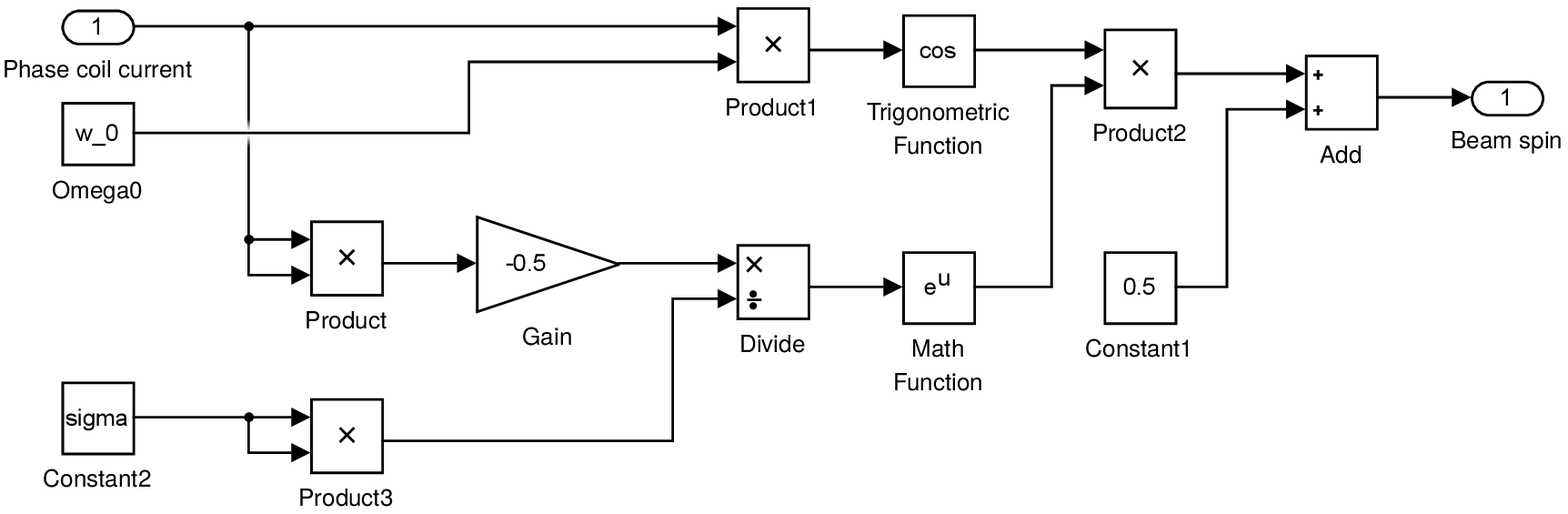}
\caption{The subsystem used to convert the phase coil current to the number of $^3$He reaching the detector according to \autoref{equ:phasecoil2}.}
\label{fig:phasecoil}
\end{figure}

\subsection{Lock-In Amplifier}
\label{sec:lockinamp}
The concept of lock-in detection can be found in many textbooks. In this work a lock-in amplifier (\autoref{fig:lockinamp}) is designed to extract one frequency component from a signal. To do this it multiplies the original signal with a sine wave of the frequency that one wants to extract. In Fourier space, this can be thought of as a convolution of the original signal with a pair of delta functions, one at the original frequency and one at minus the original frequency. It has the effect of moving the component of the original signal at the desired frequency to a constant component. A low pass filter of the form $H(s)=\frac{1}{T\cdot s+1}$ is then used so that only the D.C. part of this new signal remains, providing as an output the amplitude of the component oscillating at the desired frequency. To get the amplitude correct, a lock-in amplifier using a cosine wave is also used and the results of both are added in quadrature to account for any phase differences.

\end{document}